\documentclass[]{aa}
\usepackage{graphicx}
\usepackage{txfonts}

\begin{document}

%\inputencoding{cp1251}

\title{Precision multi-epoch astrometry with VLT cameras FORS1/2
        \thanks{Based on observations made with ESO
         telescopes at the La Silla Paranal Observatory under programme
        ID 078.C-0074 and observations made with the European Southern
        Observatory telescopes obtained from the ESO/ST-ECF Science
        Archive Facility}
        }

\author{P.F.Lazorenko 
        	\inst{1} 
  \and  M.Mayor \inst{2}
  \and  M.Dominik \inst{3} 
     \thanks{Royal Society University Research Fellow}  
  \and  F.Pepe \inst{2}
  \and  D.Segransan \inst{2}     	
    \and  S.Udry \inst{2}
   	}

   \offprints{P.F.Lazorenko}

     \institute{Main Astronomical Observatory,
          National Academy of Sciences of the Ukraine,
             Zabolotnogo 27, 03680 Kyiv-127, Ukraine
          \and  Observatoire de  Geneve, 51 Chemin des Maillettes, 
          1290 Sauverny, Switzerland   
          \and  SUPA, University of St Andrews, School of Physics \&
            Astronomy, North Haugh, St Andrews, KY16 9SS, United Kingdom
              }

   \date{}                                                            

   \abstract{ }
{ We  investigate  the astrometric performance of  the
FORS1 and FORS2 cameras of  the VLT at long time scales with emphasis
on systematic errors which normally prevent
attainning a precision better than 1~mas.
}
{The study is based on multi-epoch time series of observations
of a single sky region imaged
with a time spacing of 2--6 years at FORS1 and 1--5 months at FORS2. 
Images were processed with a technique that 
reduces atmospheric image motion, geometric distortions, and
takes into account relative displacement of reference stars in time.
}
{
We performed a detailed analysis of a random error of positions
that was shown to be dominated by
the uncertainty of the star photocenter determination. 
The component  of the random error 
corresponding to image motion was found to be caused primarily
by optical aberrations and variations of atmospheric PSF size but not
by the effect of atmospheric image motion. 
Comparison of observed and model annual/monthly epoch 
average positions yielded estimates of systematic errors for which 
temporal properties and distribution in the CCD plane are given.
At frame center, the systematic component is 
about 25~$\mu$as.  Systematic errors
are shown to be caused mainly by a combined effect
of the image asymmetry and seeing variations which therefore 
should be strongly limited to avoid generating
random and systematic errors. 
For a series of 30 images, we demonstrated presicion of 
about 50~$\mu$as stable on daily, monthly, and annual  time scales. 
Small systematic errors and a Gaussian distribution of positional
residuals at any time scale indicate that the astrometric  accuracy of the VLT
is comparable to the precision.
Relative proper motion and trigonometric parallaxes of stars in the 
center of the test field were derived 
with a precision of 20~$\mu$as yr$^{-1}$ and 40~$\mu$as 
for 17--19 mag stars. 
Therefore, distances at 1 kpc could be determinable at a 4\%  precision 
if suitably distant reference objects are in the field.
 }                                                       
{We prove that the VLT with FORS1/2 cameras 
are not subject to significant systematic errors at time scales from
a few hours to a few years providing that observations are obtained in
narrow seeing limits.
The astrometric performance of the VLT imaging cameras meets
requirements necessary for many astrophysical applications,
in particular, exoplanet studies and determination of relative
trigonometric 
distances by ensuring a high accuracy of observations,
at least 50~$\mu$as attained for image series  of 0.5 hour. 
}
{}

      \keywords{astrometry   --  atmospheric effects  --
                    instrumentation: high angular resolution --
                    planetary systems}

%  \titlerunning{High-precision astrometry on the VLT/FORS1}
%  \authorrunning{Lazorenko P.F. et al. } 

     \maketitle

\section{Introduction}
The availability of astrometric measurements 
of proper motion and parallactic displacements 
at 10-100 microarsec precision 
provide a base for many astrophysical applications,
e.g. determination of the distances to stars and their
absolute luminosity, detection of planets, microlensing
studies of the mass distribution in the Galaxy, 
dynamics of the Galaxy Center stars, etc.

The above studies imply use of very high precision astrometry, requiring
both reduction methods that are fairly insensitive to major noise
sources as well as telescopes fulfilling the precision requirements and
temporal stability. The availability of suitable instruments however
hardly matches current demand, and is in large disproportion
to envisioned future endeavors. In particular, this hinders 
programmes studying exoplanet populations by means of
astrometry, which would powerfully extend and complement efforts based on
other techniques and provide an efficient pathway towards identifying habitable
planets.

The best future prospects for high-precision astrometry can be expected
from space telescopes, but the only mission currently planned
is GAIA (Perryman et al. \cite{Perryman}). 
Achieving a single-measurement
precision below 10~$\mu$as on $V < 13$ stars, it will offer the
opportunity to discover and study several thousands of planets
(Casertano et al. \cite{Casertano}). 
However, GAIA operates as an all-sky
survey and cannot be pointed to a specific target, and its accuracy
degrades rapidly towards fainter stars 
 (Lindegren et al. \cite{Lindegren2007}).

In contrast, pointing to selected targets is possible with ground-based
telescopes, effectively measuring distances in binary star systems by
means of optical interferometry. These achieve accuracies of the order of
those of GAIA at $V > 15$. VLTI/PRIMA is  able to measure distances
between stars separated by 10\arcsec with 10~$\mu$as precision 
(Delplancke et al. \cite{Delplancke})  
with a 30 minute integration time; at a similar 
20~$\mu$as h$^{-1}$ 
precision, star separations can be measured with the Keck
Interferometer  (Boden et al. \cite{Boden}). 
Moreover, at 30\arcsec separation in pairs
of bright stars, an astrometric precision of 100~$\mu$as has been
achieved by Lane et al. (\cite{Lane2000}) 
with the Palomar Testbed Interferometer
(PTI), and Lane \& Muterspagh (\cite{Lane}) and
Muterspagh et al. (\cite{Muterspagh})
reported an accuracy of 20~$\mu$as stable over a few nights.
The availability of and access to ground-based high-precision astrometry
facilities is however extremely limited, so that no exoplanet detection
programme has yet been established.

Large ground-based monopupil telescopes 
that operates in imaging mode also can significantly contribute 
to the detection and measurement planetary systems. Unlike 
infrared interferometers, these telescopes
measure the position of a target either with reference to a single 
star or to a grid of reference stars.
For a long time, however, astrometric measurements with ground-based
imaging telescopes were believed to be
limited by about 1 mas precision due to atmospheric image motion
(Lindegren \cite{Lindegren}). 
This limitation, however, is not fundamental and rather reflects the performance
of the conventional astrometric technique.
The first high precision observations well below 1 mas were obtained by
Pravdo \& Shaklan (\cite{Pravdo}) at the Hale telescope with a $D=5$ m aperture.
In the field of 90\arcsec, they demonstrated precision
of 150~$\mu$as h$^{-1}$.
This precision was further improved by Cameron et al. (\cite{Cameron})
with the use of adaptive optics.
They reached a precision of 100~$\mu$as with a 2 minutes exposure 
and showed it to be stable over 2 months.

A detailed analysis of the process of differential measurements
affected by image motion allowed Lazorenko \& Lazorenko (\cite{Lazorenko4})
to show that the excellent results obtained by
Pravdo \& Shaklan (\cite{Pravdo})
represent the actual astrometric performance of large telescopes.
It was shown that angular observations
with very large monopupil telescopes are not atmosphere limited
due to effective averaging of phase distortions over the aperture. 
For observations in very narrow fields, atmospheric image motion 
decreases as $D^{-3/2}$ (Lazorenko \cite{Lazorenko2}) reaching below
other error components.
Also, the image motion spectrum can be further filtered
in the process of the reduction by using reference field stars
as a specific filter.
Astrometric precision greatly benefits from the use of large
$D \geq8$ m apertures. 

Besides atmospheric image motion, one can list a number of other  
systematic and random error sources that could
prevent us to reach a $100\mu$as level of the precision. 
Many sources of error depend on the telescope
and  cause long-term astrometric instability of results.
To ascertain the practical feasibility of this new astrometric method, we 
have chosen the high  performance FORS1/2  cameras 
set at the VLT with excellent seeing. 
We have undertaken several 
tests of various time scales, ranging from a few hours to  
several years.

The first test (Lazorenko  \cite{Lazorenko6}) was based on
a single four-hour series of  
FORS2 images in Galactic Bulge obtained by Moutou et al. (\cite{Moutou}).
It proved the validity of the basic concept of the new astrometric
method and an astrometric precision of 300~$\mu$as 
with a 17~s  exposure was reached.  

In the second test (Lazorenko et al.  \cite{Lazorenko7}), 
we investigated the astrometric precision of the FORS1 
camera over time scales of a few days. For this study
we used the two-epoch (2000 and 2002) image series 
(Motch  et al. \cite{Motch}), 
each epoch represented by four consecutive nights. We reached
a positional precision of $\sigma=200-300$~$\mu$as  
and detected no instrumental  systematic errors above 30~$\mu$as
at the time scales considered. The precision of a  
series of $n$ images was shown to improve as
${\sigma} /\sqrt{n}$ at least to $n = 30$, 
which corresponds to a 40--50~$\mu$as astrometric precision. 

This paper 
concludes our study of the VLT camera astrometric performance and
extends our previous short time scale results to 
intervals of 1--5 months and 2--6 years. This covers all time scales
required for
typical microlensing, exoplanet search applications, and
Galaxy kinematics studies.
In Sect.2 we outline the strategy of this study, observations, and the computation
of the star image photocenters. Astrometric reduction
based on the reduction model  (Sect.3) is described in Sect.4.
The random errors of single measurements are analyzed in Sect.5, where
we extract the image motion 
component, which proved to be of instrumental origin.
Systematic errors in epoch monthly/annual average positions,
and their spatial and temporal properties are considered in Sect.6.
Astrometric precision in terms of the Allan deviation is discussed
in Sect.7.

%\newpage

\section{Observation strategy and computation of photocenters}

As a test star field, the best choice is the field 
near the neutron star  RX J0720.4-3125 whose FORS1/UT1 images 
of 3.3$\times$3.3\arcmin angular size obtained
in Dec 2000 and Dec 2002 by Motch  et al. (\cite{Motch})
were  already used in our previous study (Lazorenko et al. \cite{Lazorenko7}).
Its uniqueness is that it has the best history of observations
available in the ESO/ST ECF Archives 
suitable for precision astrometry.  
Data are represented by 65 images obtained with the $B$ filter and
obtained with a 2 year epoch difference, 
which allows for a reduction  with no bias due to
parallax.  Also, the field is densely populated,
containing about 200 stars with a high light signal.
We repeatedly observed it in Dec 2006, 
at integer differences of years, with FORS1/UT2 (1px$=0.10$\arcsec scale) and
the same $B$ filter, thus  
comparing model predicted and observed positions 
at three annual epochs, verifying 
the very long-term astrometric stability of the VLT at 2-6 years,
and computing relative proper motions used later on 
for the reduction of FORS2 data. 
Observations were performed with the LADC optical system
(Avila et al. \cite{Avila}) that improves image quality by compensating
for the differential chromatic refraction (DCR) of the atmosphere.

The same test star field was imaged  five times at FORS2 
(1px$=0.126$\arcsec scale)
with the $R_{\rm special}$ filter with a
$T=70$~s exposure  to keep star fluxes at
approximately the same level as in the FORS1 images. 
A one month spacing between time series was chosen to match
the  typical sampling for the observation of
astrometric microlensing or of the astrometric shift of stars caused by an
orbiting planet.
The availability of FORS1 images gives us an opportunity to correct
the measured FORS2 positions for highly accurate 
relative
proper motions determined at six year time intervals.
This correction is critically important since elimination of proper
motion from FORS2 positions
decreases the number of model parameters, thus greatly improving
the reliability of the subsequent statistical analysis.
After elimination of proper motion obtained as shown, 
the positions of 5 series are reduced to a common standard frame
with a model that fits star  motion by parallax.  
Residuals of  star positions (measured minus model)
are then analyzed to detect   systematic errors
and any correlations with  time or magnitude. 
A summary of observation data is given in Table \ref{obs}.
Note the large variations of seeing which does
not favor high precision astrometry (Sect.6).

The primary goal of this study is the investigation of random and systematic
positional errors of the FORS cameras. This task 
requires a careful reduction of observations, including determination
of proper motion and parallaxes with the combined use of images 
obtained with two cameras. We aim to demonstrate that a 300$\mu$as single 
measurement precision of narrow-field astrometry translates to about 50$\mu$as
precision for a series of 30 measurements.

\begin{table}[tbh]
\caption [] {Summary of the test field observations}
\begin{tabular}{@{}l@{}ccccc@{}}
\hline
\hline
         & No. of nights &  camera         & spectr.& T     &   seeing,   \rule{0pt}{11pt}\\ 
Date     & /images       & /unit  & band   & sec   &   arcsec   \\ 
\hline
Dec 2000 & 4/40     & FORS1/UT1   &  $B$   & 620 & 0.49--0.78  \rule{0pt}{11pt}\\
Dec 2002 & 4/25     & FORS1/UT1   &  $B$   & 620 & 0.46--0.83  \\
Dec 2006 & 1/5      & FORS1/UT2   &  $B$   & 560  & 0.55--0.65 \\
Nov 2006 & 1/27     & FORS2/UT1   &  $R$   & 70  & 0.48--0.62 \\
Dec 2006 & 1/37     & FORS2/UT1   &  $R$   & 70  & 0.34--0.73 \\
Jan 2007 & 1/27     & FORS2/UT1   &  $R$   & 70  & 0.56--0.89 \\
Feb 2007 & 1/27     & FORS2/UT1   &  $R$   & 70  & 0.45--0.63 \\
Mar 2007 & 1/27     & FORS2/UT1   &  $R$   & 70  & 0.57--0.76 \\
\hline                
\multicolumn{5}{l}{Standard limits for seeing*:} & 0.47--0.78 \rule{0pt}{11pt}\\
\hline
\end{tabular}
\begin{list}{}{}
\item* Images with seeing out of these limits are affected by large random and
systematic errors in positions (Sect.6).
\end{list}
\label{obs}
\end{table}

Raw images were calibrated (debiased and flat-fielded) using
calibration master files. Star images having even 
a single saturated pixel were marked and rejected
for a loss, even small, of positional information.
Positions of star photocenters $X,Y$ were computed
with the profile fitting technique based on the 12-parameter model
specific for the VLT images  (Lazorenko \cite{Lazorenko6}).
By careful examination, we developed a three component model
that fits observed profiles to the photon noise limit.
The dominant model component that approximates the core of the
PSF (point spread function) is a relatively compact
Gaussian with width parameters $\sigma_{\rm{G}}^x$,
$\sigma_{\rm{G}}^y$ along  $x,y$ axes and
a flux $I_{\rm G}$ containing  2/3 of the
total star flux $I$. Two auxiliary
Gaussians, each one multiplied by a 
factor $x^2$ or $y^2$, are
co-centered at the dominant component and approximate
wings of the PSF. The model also takes into account the high-frequency oscillating
feature of the PSF,  computing it as a systematic discrepancy
between the model and observed star profiles. 
Deviations
between the model and observed pixel counts were found 
to be at the $\chi^2 \approx 1$ level.
Determination of  star photocenters is a very
important element of the  process because, as we show later on, 
most of the random and systematic errors occur at this phase.

The precision  $\varepsilon$ of  the star photocenter was
estimated by numerical simulation of random images
yielding an expression 
similar to that derived by Irwin (\cite{Irwin})
for a single Gaussian profile
\begin{equation}
\label{eq:phsigma}
\varepsilon =  \phi_2 \frac{ \rm FWHM}{2.34 \sqrt{I_{\rm G}}}
\sqrt{1+ \phi_1  \frac{8 \pi  \sigma_{\rm{G}}^2  I_{\rm {b}}}{ I_{\rm G}  }}.
\end{equation}
This equation is valid in a much wider range 
of fluxes, seeing, and background signal $I_{\rm {b}}$ as compared to our 
former expression (Lazorenko et al. \cite{Lazorenko7}).
Here fluxes are given in electrons, $\sigma_{\rm{G}}$ is expressed
in pixels, $ \phi_1=0.820$ and
$ \phi_2= 1+0.15 ( \sigma_{\rm{G}}-1.5)^2$ are empiric factors,
and ${\rm FWHM} = 3.10 \sigma_{\rm{G}}$ is a
relation valid for the VLT images. Due to the complex star profiles,
coefficients $\phi_1$ and $\phi_2$ are not units, in  which case
Eq.(\ref{eq:phsigma}) transforms to the expression given
by Irwin (\cite{Irwin}).

Computed model parameters were analyzed to detect and reject
non-standardly shaped images  indicating  computation errors and
actual image defects caused by blending, cosmic rays, etc. 
All images with model parameters and $\chi^2$ 
exceeding some deliberately set 
thresholds were discarded. Thresholds  were
chosen so that the frequency of rejections
was about 1\%  for bright images. 
At this fixed threshold, the number of rejections
gradually increased with magnitude, 
reaching 10--25\% for faint images,
which  are more sensitive to the background irregularities.
In contrast, filtration based on  $\chi^2$ caused excessive rejection
of the brightest images, since the accuracy of the model profile
is insufficient at high light signals and becomes  
comparable to the statistical fluctuations of counts.
This gives rise to a $\chi^2$ with subsequent false rejection of measurements. 

\begin{figure}[htb]
\centering
\resizebox{\hsize}{!}{\includegraphics*[]{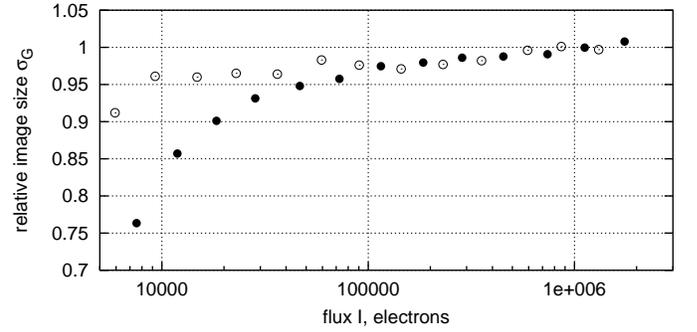}}
\caption {Ratio $ \sigma_{\rm{G}}  / \sigma_{\rm{init}}$
of the mean star image size in the selected star sample with good images
to its value in the initial sample
as a function of flux  for FORS1 
(filled circles) and  FORS2 (open circles). 
}
\label{soni}
\end{figure}

Most often, discarded faint star images have excessive size.
This produced a selective effect seen as 
a systematic dependence
of image parameters on flux $I$.
A difference between the mean image size parameter $\sigma_{\rm{G}}$ 
in the filted star sample and
its mean value $\sigma_{\rm{init}}$ in the initial star sample is typical.
The systematic dependence of the 
ratio $ \sigma_{\rm{G}}  / \sigma_{\rm{init}}$
on flux is shown in  Fig.\ref{soni}.
While no difference is seen for bright images, at the faint end 
the size
$\sigma_{\rm{G}}$ of stars selected for further processing is
systematically 10--30\%  smaller.
For FORS1 images the effect is stronger due to the larger pixel
scale (lower signal to noise ratio) and  many cosmic rays occured over the long
integration time.
The selection described here induces a similar
dependence  of the centroiding error  on flux (Sect.5). 

Further astrometric reduction 
revealed  that some stars show a significant correlation
between  model minus observed  residuals 
of positions and variations of seeing. 
This effect, detected primarily for relatively close star pairs, 
is due to the asymmetry of images  caused by the light from 
the nearby star (see Sect.6). About 
1\% of measurements subject to this effect were rejected.

%\newpage
%yyy
%\newpage
\section{Astrometric reduction model}

$\omega(R)$ is a sample of $i=0,1,2 \ldots N$  stars 
imaged  $ m=0, \ldots M$ fold
in the sky area of  angular radius $R_{}$
centered on a target star which we denote with a subscript
$i=0$. 
In general, $\omega(R)$ may represent only a portion of the
complete sample of
stars  $\Omega$ seen in  the telescope FoV. 
Given the measured  star centroids  $X_{im}$, $Y_{im}$,
we derive,  on each CCD image, the differential position of the target,
its 
relative parallax, proper motion, and deviations $V_{im}$ from
the model that may hide the  astrometric signal (e.g. planetary
signature). 
These quantities are not measured directly and are rather estimates of
model parameters found in a certain {\it reference system} set by
the reduction model; therefore, they depend on it.
The image $m=0$ sets the zero-point
of positions,  a grid of reference stars in this image defines
the {\it reference frame}. For processing therefore we use relative CCD
positions $x_{im}=X_{im}-X_{i0}$, $y_{im}=Y_{im}-Y_{i0}$.

The certain difficulty for astrometric processing is brought by 
the instability of the reference frame in time due to 
the atmospheric differential chromatism (e.g. Monet et al. \cite{Monet};
Pravdo \& Shaklan \cite{Pravdo}; Lazorenko \cite{Lazorenko6};
Lazorenko et al. \cite{Lazorenko7}), 
variable geometric distortion, proper motions, etc. 
In our
previous study (Lazorenko et al. \cite{Lazorenko7}), we developed 
a model that correctly handles this problem 
and  ensures
a solution in a uniform system with no distinction between target
and reference stars. Here we propose 
a more general solution that 
allows an easy readjustment of the system of model parameters and 
of deviations $V_{im}$  in a way that is optimal for a particular study.
The model deals with atmospheric image motion (Sect.3.1), geometric
distortions  (Sect.3.2), and instability of the reference frame in time (Sect.3.3).
We emphasize that all the data derived from the differential
reduction (proper motions, parallaxes, chromatic parameters, etc.)
are intrinsically {\it {relative}} (not absolute). This point
is discussed  in detail in Sect.3.5.

%%%%%%%%%%%%%%%%%%%%%%%%%%%%%%%%

\subsection{Atmospheric image motion}
The variance of the atmospheric image motion 
in positions measured in narrow fields is  given by
Lindegren's (\cite{Lindegren}) expression  
\begin {equation}                               
\label{eq:lind}
\delta^2 \sim   {(hR/D)}^{4/3} {(hR)}^{2/3} {T}^{-1}
\end{equation}
where $h$ is the altitude 
of the atmospheric turbulent layer generating the image motion
and $R$ is a star configuration angular size. For a binary star, $R$ is the
star separation. Eq.(\ref{eq:lind}) 
refers to the very narrow angle observations defined by condition
\begin {equation}                               
\label{eq:nangl}
hR < 0.5 D,
\end{equation}
otherwise we have the  much worse
$\delta^2 \sim {(hR)}^{2/3} {T}^{-1}$.  
Eq.(\ref{eq:lind}) predicts a weak improvement of $\delta$ with $D$
thus limiting ground-based observations to a few milliarcsec precision
at reasonable $R$ and $T$. For a fictitious case of {\it symmetric}
continuous distribution of reference stars around a target in a circle
of a radius $R$, Lindegren obtained
$\delta^2 \sim   {(hR/D)}^{2} {(hR)}^{2/3} {T}^{-1}$ with
a stronger dependence on $D$. However,
no way to  practically implement
potentially useful symmetric distributions was found, thus current
estimates of $\delta$ are based on Eq.(\ref{eq:lind}) with
milliarcsec limitation.
                  
We have shown 
(Lazorenko \cite{Lazorenko2}; Lazorenko \& Lazorenko \cite{Lazorenko4})
that
\begin{itemize}
\item  any arbitrary {\it discrete} reference star distribution, 
       of at least three stars, can be  symmetrized;
\item  symmetrization is always implemented by a standard plate reduction
            with a linear or more complex model;
\item  $\delta^2 \sim   {(hR/D)}^{3} {(hR)}^{2/3} {T}^{-1}$, 
       which suggests  faster  improvement of $\delta$ 
       with $D$ in comparison to Lindegren's
       prediction for symmetric continuous distributions;
\item   use of a {\it large} $D \geq 8$~m is required to meet condition (\ref{eq:nangl}) 
        for high stratospheric layers.
\end{itemize}

Here we briefly summarize the concept of image motion
reduction based on the spectral description of this process.
We have shown that
the spectral power density $G (q)$ of differential image motion
in the domain of spatial frequencies $q$ related to the turbulent layer
is the product of two factors.
The first factor $F'(q)$ depends on  $D$,  
the altitude $h$ and 
properties of the atmospheric turbulent layer generating the image motion,
and  exposure time $T $.
The second factor $F''$ depends only on the way we define the 
"differential position"  and on the geometry
of reference star distribution relative to the target. This factor
is expanded in a series of even powers of $q$. Hence $G(q)$ has
a simple modal structure
$G(q)=  F'(q) \sum_{s=1}^{\infty} q^{2s}F''_{2s}(x_0,y_0,x_i,y_i)$
with terms $F''_{2s}$ dependent on the distribution of stars on the sky only.
Integration of the power density 
over $q$ yields the  variance of differential image motion 
\begin {equation}                               
\label{eq:sadef}
\delta^2 =  {\left (
                \frac{h_{\rm }R}{D}
         \right )}^3   \frac{{(h_{\rm }R)}^{2/3}}{T}
        \sum_{s=1}^{\infty}H_{2s}F''_{2s}(x_0,y_0,x_i,y_i)
\end{equation}
that inherits the initial modal structure of $G(q)$, and
$H_{2s}$ are modal coefficients. 
Of course, the  actual value of $\delta^2$ is the  sum over all
turbulent layers with $h$ ranging from a few to 25 km. 
The factor $(hR/D)^3$ shows that large telescopes 
easily suppress even high stratospheric turbulence. 
Eq.(\ref{eq:sadef}) 
however is valid only for  observations at narrow angles (\ref{eq:nangl}). 
In the case of the VLT, it holds only approximately at
$R \leq$ 0.5--1\arcmin, which causes a problem of
a lack of reference stars.
With $D<8$~m apertures, only low-layer turbulence is well reduced,
resulting in a moderate suppression of the image motion.

For optical interferometers, the atmospheric noise
decreases as $d^{-2/3}$ (Shao \& Colavita \cite{Shao}) and 
phase fluctuations are mitigated  by probing the  difference of phase
at ends of the instrument long baseline $d$. For  monopupil telescopes,
a fast  decrease of image motion occurs in another way, by
averaging the turbulent phase fluctuations over the aperture. 
The efficiency of phase averaging depends on 
the symmetry of the star configuration, which requires availability
of the {\it grid} of reference stars. We emphasize that, unlike
optical interferometers, monopupil telescopes are not adapted for
precision measurement of the angular offset between a pair of stars 
due to the intrinsic
asymmetry of this star configuration, for which $\delta^2$ follows
dependence (\ref{eq:lind}).

From  Eq.(\ref{eq:sadef}) it follows that 
$\delta^2$ can be reduced by removing (filtering)
the several first most significant modes $s=1,2 \ldots$ up to some
optional $k/2$ spectral mode.
The residual variance  
$\delta^2 $
obtained in this way
depends on the first high  $k=2s$ ($k$ is even integer) active mode 
of the image motion spectrum and is therefore of a comparatively small magnitude.
For the VLT, the  gain in  $\delta^2$ is a factor of 100 and 
more. The above possibility follows from the next considerations:
the  relative position of the target in image $m$ along the $x$ axis
(here and farther on we omit similar expressions for $y$) 
is  defined by the quantity 
\begin {equation}                               
\label{eq:xdef}
V_{0m}=  {\sum \limits _{i \in \omega}}'  a_{0i}(x_{0}-x_{i})=
       x_{0} - {\sum \limits _{i \in \omega}}' a_{0i}x_{i}, 
\end{equation}
where prime indicates that index $i=0$ is omitted. 
Coefficients $a_{0i}$ meet 
the normalizing condition $\sum'  a_{0i}=1$ and are specified below.
Image motion in $V$ is described by Eq.(\ref{eq:sadef}) with 
$ F''_{2s} = \sum_{w=w_1}^{w_2}g_{ws}(f_{0w}-{\sum}'_i a_{0i}f_{iw})^2$
where  $g_{ws}$ are constants and $f_{iw}$ are values of functions
$f_{w}$ for a star $i$. Here  $f_{w}$  are
basic functions numbered with an index $w=1,2 \ldots$ defined
in Cartesian coordinates of the reference frame  
and formed with successive integer powers of the coordinates:
$1,\,x,\,y,\,x^2,\,x y,\,y^2 \ldots$. 
Thus  $f_{i1}=1, \, f_{i2}=x_i, \, f_{i3}=y_i, \, f_{i4}=x_i^2 \ldots$ 
for $i$ star. 
The expression for $ F''_{2s}$ that refers to some mode $s$
involves only polynomials of $x,y$ with a sum of powers equal to $s$.
Indices $w$ run from $w_1=s(s+1)/2+1$ to $w_2=(s+1)(s+2)/2$.
The quadratic structure of the expression for $F''_{2s}$ implies that 
the $s$ mode is zero when coefficients $a_{0i}$ meet conditions
\begin {equation}                               
\label{eq:adef}
{\sum \limits _{i \in \omega}}'  a_{0i}f_{iw} =f_{0w}
\end{equation}
for each $w= w_1 \ldots w_2 $ basic function. Evidently,
all modes up to $s=k/2 -1$ vanish if $a_{0i}$ satisfy equations 
(\ref{eq:adef}) for each $s<k/2$.
For that reason, $ a_{0i}$ are found as a solution 
of a linear system of
\begin {equation}                               
\label{eq:n}
  N'= k(k+2)/8
\end{equation}  
equations  (\ref{eq:adef}). 
Solution of this redundant system (because usually $N>N'$)  is found
with a supplementary condition 
\begin {equation}                               
\label{eq:cond}
{\sum \limits _{i \in \omega}}'  a_{0i}^2 \varepsilon_{i}^{2} = {\rm min}
\end{equation}
set on the variance $\Delta_{\rm rf}^2=
\sum a_{0i}^2 \varepsilon_{i}^{2}$ 
of the second item in (\ref{eq:xdef}) caused by centroiding errors
for reference stars.   For dense sky star areas the  approximate expression
\begin {equation}                               
\label{eq:rff}
 \Delta_{\rm rf} \approx  \frac{ {\rm FWHM}}{2.36 R \sqrt{\pi I'}}
            \frac{k}{4} 
\end{equation}
is valid where $I'$ is the  light flux per unit area coming from
bright reference stars.

Thus, the quantity $V$  defined by (\ref{eq:xdef}) is free from the  first 
modes of the image motion spectrum untill $k/2$ providing that
$ a_{0i}$ confirm to Eqs.(\ref{eq:adef},\ref{eq:cond}). 
The variance of  $V$ is
\begin {equation}                               
\label{eq:var}
\sigma^2_0 = \varepsilon_{0}^{2} + \delta^2_0 + \Delta_{\rm rf}^2 .
\end{equation}
Because $\delta^2 \sim  R^{11/3}$ due to (\ref{eq:sadef}),
and  $ \Delta_{\rm rf}^2 \sim R^{-2}$ 
according to  Eq.(\ref{eq:rff}), the value of $\sigma^2$ is minimum at
\begin {equation}                               
\label{eq:opt}
 \delta_0 = \Delta_{\rm rf},
\end{equation}
which is reached at some  
optimal size $R=R_{\rm opt}$ (see Table \ref{RR}) of  
the reference frame $\omega(R)$.

%%%%%%%%%%%%%%%%%%%%%%%%%%%%%%%%%%%%%%%%%%%

\subsection{Single plate reduction}
A standard plate reduction produces effects equivalent to symmetrization
of the reference frame
(Lazorenko \& Lazorenko \cite{Lazorenko4}). 
Really, the basic equation of the plate model, in vector representation, 
is 
\begin {equation}                               
\label{eq:f1}
        \vec {fc=x}
\end{equation}
where $\vec {x}$ is  a $1 \times (N+1)$ vector of $x$ positions 
(including target),
$\vec {f}$ is  a matrix formed by vectors $f_{iw}$,
and $\vec{ c}$ is a $1 \times N'$ vector of $N'$ model parameters $c_w$.
We require that 
$N'$ takes only those discrete values which are
defined by (\ref{eq:n}) for some optional $k$
($k=4$ and $N'=3$ corresponds to the linear model,
$k=6$ and $N'=6$ refer to the model that includes quadratic powers of 
$x,y$, etc).
The least square solution of (\ref{eq:f1}) is 
$\vec {c=F^{-1}f^{\rm T}Px}$ where 
$\vec {F=f^{\rm T}Pf}$ is the normal $N' \times N'$ matrix and
$\vec {P}$  is the diagonal matrix of weights $P$ 
%which 
%are 
assigned differently for the target and reference stars.  
To comply with the image motion reduction procedure, we set 
\begin {equation}
\begin{array}{ll}                               
\label{eq:pp}
  P_0=0 &\; {\rm for \; target}     \\
  P_i=\tilde{\sigma}^{-2} &\; {\rm for \; reference \; stars}     
\end{array}
\end{equation}
considering that 
$\tilde{\sigma}^2= \varepsilon_{}^{2} + \delta^2$
is the effective variance of  measurements. 
The best estimate of the vector  $\vec {x}$ is 
$\vec{ {\hat x}=ax}$  where $\vec {a=fF^{-1}f^{\rm T}P}$
is a projective $(N+1) \times (N+1)$ matrix that maps the matrix $\vec {f}$ 
to itself:
\begin {equation}                               
\label{eq:pro}
 \vec {af=f}      .
\end{equation}
Residuals of the least square fit is the  vector $\vec {V= x - \hat x}$, or
\begin {equation}                               
\label{eq:v}
        \vec {V= x - ax}
\end{equation}
with the  property
\begin {equation}                               
\label{eq:propv}
        \vec {V^{\rm T}Pf=0} .
\end{equation}
The covariance 
matrix of  $\vec {V}$ is 
$\vec {B= \{VV^{\rm T}\}}$ where  curly brackets designate
mathematical expectation. Using  Eq.(\ref{eq:v})
and considering that $\vec { \{xx^{\rm T}\}=\tilde{\sigma}^{2}}$, we find 
$\vec {B= \tilde{\sigma}^{2}- P^{-1}a^{\rm T}- aP^{-1} +aP^{-1}a^{\rm T} }$
where $\vec {\tilde{\sigma}^{2}}$ is a diagonal matrix of elements
$\tilde{\sigma}^{2}$.
Diagonal $ii$ elements of $\vec {B}$ are a variance ${\sigma}^{2}_i$ of the
residuals $V_i$.
For target ($P_0 = 0$),
all elements in column $i=0$  of $\vec {a}$ are zero. Hence
${\sigma}^{2}_0  = \varepsilon_{0}^2 + \delta_{0}^2+ \Delta_{\rm rf}^2$
which is equivalent to  Eq.(\ref{eq:var}) where 
\begin {equation}                               
\label{eq:dr0}
\Delta_{\rm rf}^2=(\vec { aP^{-1}a^{\rm T} })_{00}=
(\vec {fF^{-1}f^{\rm T}})_{00}
\end{equation}
defines the variance of reference frame component $\vec{fc}$ at the
location of the target.
For reference star $i$ we come to a different expression
\begin {equation}                               
\label{eq:dr}
 {\sigma}^{2}_i = \varepsilon_{i}^2 + \delta_{i}^2 -  (\vec { aP^{-1} })_{ii}
   = \varepsilon_{i}^2 + \delta_{i}^2 - (\vec {fF^{-1}f^{\rm T}})_{ii}.
\end{equation}
The last term is the noise from the reference frame and 
is  a
composition of two components, noise from the star $i$ itself with
the variance $\tilde{\sigma}^{2}_i$, and nearby star grid noise
with the variance  $\Delta_{\rm rf}^{2}$ defined by
Eq.(\ref{eq:dr0}) at the location of star $i$. Adding their inverse
as weights, we find  $(\vec {fF^{-1}f^{\rm T}})_{ii}^{-1}=
\tilde{\sigma}^{-2}_i+\Delta_{\rm rf}^{-2}$. Hence
\begin {equation}                               
\label{eq:dr2}
  {\sigma}^{2}_i = \tilde{\sigma}^{2}_i - \Delta_{\rm rf}^{2} \tilde{\sigma}^{2}_i/
  (\Delta_{\rm rf}^{2}+\tilde{\sigma}^{2}_i).
\end{equation}

Eqs.(\ref{eq:v}),  (\ref{eq:pro}) of the
plate model correspond to Eqs.(\ref{eq:xdef}), (\ref{eq:adef})
of the image motion filtration and  (\ref{eq:cond}) is the 
least square condition. Therefore both methods are equivalent.
However, we imply that the model (\ref{eq:f1}) should include
a sequence of all
basic functions $f_w$ with {\it no omission} and 
at least $k=4$ (linear plate solution) or above is chosen. 
The use of higher-order  models 
results in better filtration of image motion,
though, as follows from Eq.(\ref{eq:rff}), it increases  $\Delta_{\rm rf}$.
In the  special case of  $k=2$, differential positions are measured
relative to the centroid of the reference stars; 
this should be avoided since it corrupts
the  symmetry of the  reference configuration and greatly amplifies image motion
to $\delta^2 \sim {(hR/D)}^{4/3} {(hR)}^{2/3} {T}^{-1}$ set by 
(\ref{eq:lind}).

%%%%%%%%%%%%%%%%%%%%%%%%%%%%%%%%%%%%%%%%%%%%%%%%%%%%%%
        
\subsection{Multi-plate reduction}
A single plate model (\ref{eq:f1}) is easily extended to the case of
multiple  $m=1, \ldots M$ images. For this purpose we 
specify, for any  star $i$,
a set of  $s=1, \ldots S$ model parameters $\xi_{is}$ which are
zero-points, 
relative proper motion $\mu^{x}$ and $\mu^{y}$, 
relative parallax  $\pi$, 
etc. Thus the position  $x_{im}$ of any star $i$ in image $m$ is modelled by
$x_{im}= \sum_{w=1}^{N'} f_{iw} c_{wm} + \sum_{s=1}^{S} \xi_{is} \nu_{sm} $
where $\nu_{sm}$ are functions of time (of image number) coupled to $\xi_{is}$
and $ c_{wm}$ are model
parameters $ c_{w}$ in image $m$.   Using matrices similar to
the above quantities, we reach
\begin {equation}                               
\label{eq:mod2}
   \vec {fc + \xi \nu = x}
\end{equation}
in a concatenated space formed by two types of basic functions, 
$ {f}$ and $ {\nu}$, related to the spatial coordinates or
to time respectively. A similar expression is written for $y$,
which requires introduction of the corresponding $\vec c^{(y)} $ matrix.

Eq.(\ref{eq:mod2}) is a set of $(N+1) \times M$ equations solvable by the
least square fit with respect to $N' \times M$ parameters  $c_{wm}$ and
$(N+1) \times S$ parameters $\xi_{is}$. Also, for each image $m$, we introduce 
an $(N+1) \times (N+1)$ weight matrix $\vec {P(m)}$
with elements $\tilde{\sigma}^{-2}_{im}$  used for the reduction in $x$,$y$
space and treated like
the matrix $\vec {P}$ of Sect.3.2. 
Another $M \times M$ diagonal matrix $\vec {P(i)}$ related to 
a star $i$ takes into account the change in time of the 
residuals $\vec {(x-fc)}_{im}$. Diagonal  elements 
of $\vec {P(i)}$ are equal to ${\sigma}^{-2}_{im}$
defined by Eq.(\ref{eq:var}) or Eq.(\ref{eq:dr2}) depending on the star type. 
When star $i$ measurements are
unavailable at image $m$, the corresponding elements of matrices
$\vec {P(i)}$ and $\vec {P(m)}$  are put to zero.

Direct solution of  system (\ref{eq:mod2}) is, however,
impossible due to the  ambiguity between $\vec{ \xi}$ and $\vec{ c}$.
If some component $s$ of parameters $\xi_{is}$ (for example, proper
motion) systematically changes across the CCD, this change can be 
fitted by a polynomial and thus is not resolvable from a change in $c_{wm}$.
And vice versa,
a change in time of some component $w$ of $c_{wm}$ produces an effect
that is similar to a change in $\xi_{is}$. Therefore
Eq.(\ref{eq:mod2}) is solved under $S \times N'$ conditions
\begin {equation}                               
\label{eq:cond2}
   \vec {\xi^{\rm T} {\bar P} f = 0}
\end{equation}
where $\vec {\bar P}$ is the $N \times N$ diagonal matrix 
of weights ${\bar P}_i$. For reference stars, ${\bar P}_i$ are
{\it arbitrary} (e.g. average of $P_{im}$ over all measurements) while
${\bar P}_0=0$ for the target.
The least square estimates $\vec {c}$  and $\vec {\xi}$ of Eq.(\ref{eq:mod2})
are found from the system of equations 
\begin {equation}                               
\begin{array}{ll}
\label{eq:sol2}
   \vec {c=  F^{-1}(m) f^{\rm T} {P(m)} (x - \xi \nu)} \\
   \vec{ \xi = (x - fc) P(i) \nu^{\rm T} N^{-1}(i)  }
\end{array}
\end{equation}
solved iteratively using reference stars only with no contribution
from the target due to zero weights $P_{0m}$ and  ${\bar P}_0$. 
Above,  $\vec{N(i)=\nu P(i) \nu^{\rm T}}$
is a normal matrix for a star $i$ in vector ${\nu}$ space and
$\vec{F^{-1}(m)}$ is a matrix $\vec{F^{-1}}$ related to image $m$. 
For simplicity, in Eq.(\ref{eq:sol2}) we omit expressions
related to $y$.

A simple, non-iterative 
solution of Eq.(\ref{eq:sol2}) exists, requiring only
that each star measurement is available at each image,
at constant flux and seeing conditions.
In this case $P_i={\rm const}$ and therefore 
$\vec {c=  F^{-1}(m) f^{\rm T} {\bar P}x}$ and 
$ \vec{ \xi = (x - ax) {\bar P} \nu^{\rm T} N^{-1}(i)  }$
providing  that ${\bar P}_i=P_i$ is set.

Given solution  $\vec {c}$  and $\vec {\xi}$, we derive residuals 
\begin {equation}                               
\label{eq:v2}
\vec {V=x-fc- \xi \nu}
\end{equation}
which are orthogonal to the basic vectors $\vec{f}$ (at each $m$) and $\vec{\nu}$
(for each $i$): 
\begin {equation}
\begin{array}{lr}                               
\label{eq:ort}
 \vec {V^{\rm T}  P(m) f = 0}; &  \qquad \vec {V  P(i) \nu^{\rm T} = 0.}
\end{array}
\end{equation}

Putting solution $\vec {c}$ derived from measurements
of reference stars only into the second equation  of (\ref{eq:sol2}), 
we find parameters $ { \xi_{0,s}}$ of the target.
Hence, the variance of $V$ at image $m$ is 
\begin {equation}                               
\begin{array}{l}                               
\label{eq:vv}
   \sigma^2_{0m} =   \varepsilon_{0}^2 + \delta_{0}^2 +\Delta_{\rm rf}^{2}
   - [\vec {\nu^{\rm T} N^{-1}(0) \nu } ]_{mm} 
    \quad  {\rm for \; target} \; i=0 \\

   \sigma^2_{im} =  \tilde{\sigma}^{2}_i - 
         \Delta_{\rm rf}^{2} \tilde{\sigma}^{2}_i/
  (\Delta_{\rm rf}^{2}+\tilde{\sigma}^{2}_i)
   - [\vec {\nu^{\rm T} N^{-1}(i) \nu } ]_{mm}
    \quad  {\rm ref. \; star} \; i.   
\end{array}
\end{equation}

%%%%%%%%%%%%%%%%%%%%%%%%%%%%%%%%%%%%%

\subsection{Reference frame quality}
\begin{figure}[htb]
\resizebox{\hsize}{!}{\includegraphics*[bb = 58 54 308 170]{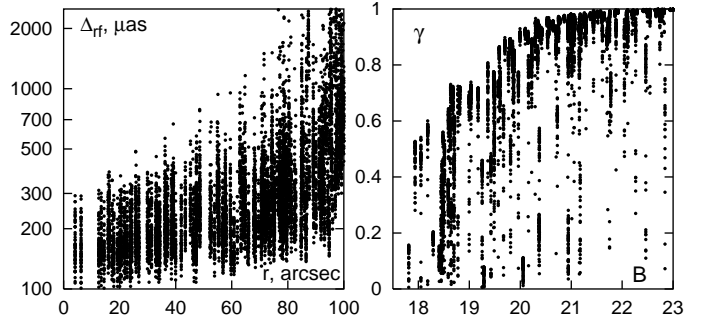}}
\caption {Characteristics of reference frame: 
error $\Delta_{\rm rf}$ as a function 
of distance $r$ of the target from the field center ({\it left panel}) and 
the quality factor $\gamma$ (dependence on 
magnitude) ({\it right panel}) for each star and for each of 70 FORS1 images.
The vertical scatter of dots reflects variation of seeing and background
in time.
Reduction was performed in the standard way ($P_0=0$,
$ {\bar P}_0=0$) with $k=10$ (model with fourth powers of $x,y$) 
and optimal $R=1.5$\arcmin.
}                                          
\label{gam}
\end{figure}

In very dense reference frames,  astrometric precision is limited only 
by errors
of the photocenter determination if
image motion is well reduced. For sparsely populated reference
frames, the noise $\Delta_{\rm rf}$  degrades the astrometric precision. 
This takes place even in
our case of  relatively low galactic latitude, -8\degr.
Fig.\ref{gam}  shows the distribution of 
$\Delta_{\rm rf}$ computed for each star (processed as a target) and each
FORS1   image as a function of star distance $r$ from the frame center.
The vertical scatter of  dots for each star is rather large and is
caused by variations of seeing and background.  
At frame center,  $\Delta_{\rm rf}$  is minimum, about $150$~$\mu$as in average
observing conditions. Although small, it is 
comparable to the centroiding error,  $\varepsilon \approx 300$~$\mu$as,  
of the brightest targets.
A good indicator of the ability of the reference frame  to keep 
the output error of positions near to 
the precision of  image centroiding is the  quantity 
\begin {equation}                               
\label{eq:gam}
\gamma=\frac{\varepsilon_{}^{2}}{\varepsilon_{}^{2}+
                  \Delta_{\rm rf}^{2}}
\end{equation}
which depends on the star location in the frame, its brightness, and
in extreme cases of $\varepsilon_{}^{2} / \Delta_{\rm rf}^{2}$  ratio, it 
varies from zero  to a unit value. In terms of $\gamma$,
the dependence of $\sigma^2$ on the reference frame noise  given by 
Eqs.(\ref{eq:var}) and (\ref{eq:dr2}) is
\begin {equation}                               
\begin{array}{l}
\label{eq:g1}
 \sigma^2_0  = \varepsilon_{0}^2 / \gamma, \qquad P_0=0 \; {\rm (target)} \\
 \sigma^2_i  = \gamma \varepsilon_{i}^2,  \qquad  P_i>0 \; {\rm (ref. \; star)}
\end{array}
\end{equation}
where for simplicity we assumed $\delta=0$.
Although written for a single image only, the 
above equations emphasize the problem caused by sparse 
reference frames with small $\gamma$. 
Note that a decrease in $\sigma^2_i$ with $\gamma$
for reference stars
does not improve the signal-to-noise ratio (see Sect.3.5).

%%%%%%%%%%%%%%%%%%%%%%%%%%%%%%

Fig.\ref{gam} shows  distribution of $\gamma$ as a function
of magnitude for each star processed as a target. 
The dependence has a specific decline  at the bright end, to $\gamma \approx 0.2$--0.6.
For bright targets $\gamma \approx 0.5 $  even at field center 
due to the  limited number of reference stars.
It follows that for small $\gamma$,
the resulting variance $ \sigma^2_0$  significantly exceeds the
centroiding error $\varepsilon_{0}^2$ since  $ \sigma^2_0 \sim 1/ \gamma  $.
In this respect,  $ {\gamma}$ is a factor that 
specifies the quality of the reference frame. 

%%%%%%%%%%%%%%%%%%%%%%%%%%%%%%%%%%%%%%%%
\subsection{System of output data}

Due to the differential reduction, the computed parameters
 $\xi $ and positional residuals $V$ are { \it relative}.
Weights ${\bar P}_i$  define
the {\it system} of parameters $\xi $ of both target and reference stars.
It follows from Eq.(\ref{eq:cond2}) which
allows interpretation  of
$\vec {\xi}$ as the residual of the least square fit of some
absolute  parameters $\vec {\xi}_{\rm abs}$ 
by basic functions $f$.
Therefore what we measure are not $\vec {\xi}_{\rm abs}$ but
relative values
\begin {equation}                               
\label{eq:syst}                 
 \vec {\xi = 
       {\xi}_{\rm abs}-{\bar a} {\xi}_{\rm abs}}
\end{equation}
where  $\vec { {\bar a}}$ is a projective matrix defined similarly to 
$\vec { { a}}$ but with weights ${\bar P}$ put instead of ${P}$,
and $\vec {{\bar a} {\xi}_{\rm abs}}$  is a least-square polynomial fit
of $\vec {{\xi}_{\rm abs}}$ over reference stars in  $\omega$.
Thus the reference frame $\omega$ and system
weights  ${\bar P}_i$ fully define  rule 
(\ref{eq:syst}) of  $\vec {\xi}_{\rm abs}$
transformation to  $\vec {\xi}$ at $k$ (or $N'$).
Recall that weights
${\bar P}_i$ for reference stars 
are arbitrary and thus so is the transformation 
(\ref{eq:syst}). From general considerations  we assume that
${\bar P}_i$ are equal to ${ P}_{im}$ averaged over $m$, and ${\bar P}_0=0$.

Noting the  similar structure of Eqs.(\ref{eq:cond2}) and (\ref{eq:ort})
(the first equation),  
we can apply the above considerations to 
the  system of residuals $V$ and  find that it is 
defined by weights  ${ P}_{im}$. 
Residuals $V$ are related to some "absolute" residuals ${V}_{\rm abs}$ by
the expression of (\ref{eq:syst}) as
\begin {equation}                               
\label{eq:systV}                 
 \vec {V = 
       {V}_{\rm abs}-{a} {V}_{\rm abs}}.
\end{equation}
There is however an essential difference in treating Eqs.(\ref{eq:syst}) and
(\ref{eq:systV}). The variance of $ {V}_{\rm abs}$ 
depends primarily on the centroiding error
while the scatter of $ {\xi}_{\rm abs}$ 
is caused by the actual dispersion of star parameters (e.g. proper motions)
within $\omega$ and can largely exceed random errors.
Therefore, comparison of
$\xi$ values computed with different $k$, $R$ or $\omega$,
will show  a divergence  dependent 
on the particular spatial distribution and the  dispersion of 
$ {\xi}_{\rm abs}$.  For example, 
the estimates of  parallaxes computed in this test field for the same target,
but with different $k$ and $R_{\rm opt}$, are scattered 
with a standard deviation of about
$\pm100$--200 $\mu$as independent of the target brightness, which exceeds
the errors of parallax determination. 
This is quite normal and mirrors the change of reference system.
Therefore, estimates of $\xi$ obtained in different systems not can be
merged into a single system, which will produce meaningless result.
Unlike this, the change of reference system affects $V$ very much 
less (by an order) which proves their merging into a single "system" (Sect.4.3).
 	                   
System weights ${P}_{im}$ and ${\bar P}_i$  essentially affect
output residuals $V$ and model parameters ${\xi}$, which is better
analyzed 
from the point of view of signal detection. 
$z\varepsilon_{}$ 
is a signal in $x$ that generates some response $z'$ in $V$. 
With regard to the target,
the amplitude of $z'$, according to Eq.(\ref{eq:v}), is
$z'=z\varepsilon_{0}$ 
since $a_{i0}=0$. With Eq.(\ref{eq:g1}) defining
the variance of $V_0$, we find that
the signal-to-noise ratio $\eta=z' / {\sigma_0}$
is $\eta= z\sqrt{\gamma}$. Thus, while the measured
signal $z$ in $V_0$ is independent of properties of the reference field,
the signal-to-noise degrades at low $\gamma$,  primarily 
for bright targets. Now consider the reference star $i$. In 
this case $z'=(1-a_{ii})z\varepsilon_{i}$ according to Eq.(\ref{eq:v}).
From Eq.(\ref{eq:dr}) and the definition of $\gamma$
we find  $\gamma=1-a_{ii}$ hence $z'=\gamma z\varepsilon_{i}$
and $\eta= z  \sqrt{\gamma}$. 
We conclude that the signal-to-noise ratio is equal for either type of stars,
but the best 100\% response $z'$ in $V$ is detected for targets. For reference
stars, the signal decreases as $\sim \gamma$, 
especially significant for bright
stars.

For some specific studies dealing with a full sample of stars
(kinematics of open cluster members),
uniformity  of the system  of  output model parameters
$\vec {\xi}$  is much desired.
In this case, the best way is to process
all stars as reference objects ($P_i \neq 0$, ${\bar P}_i  \neq 0$).
The model solution $\vec {\xi}$ is then related to  $\vec {\xi}_{\rm abs}$
via Eq.(\ref{eq:syst}). With respect to  proper motions, this transformation is
$\mu_{i}={\bar \gamma}\mu_{{\rm abs},{i}}-
{\sum'_{j \in \omega}} a_{ij}\mu_{{\rm abs},{i}}$ where
${\bar \gamma}=1-{\bar a}_{ii} < 1$ corresponds to the system of weights 
${\bar P}_i$.

Untill now we have discussed reduction with reference to a star grid
within a single isolated circular area
$\omega(R)$  disregarding other stars in the FoV.
In our previous study
(Lazorenko et al. \cite{Lazorenko7}), we considered
reduction with multiple overlapping
reference subframes $\omega(i,R)$ 
each centered at each $i$ star seen in the FoV. 
In this approach, the star $i$ is processed, at first, as a target ($P_i=0$)
measured with reference to its own local subframe $\omega(i,R)$.
At the same time, this star is 
a reference object  ($P_i \neq 0$) for adjacent subframes.
The solution initially related to local frames $\omega(i,R)$ 
is iteratively expanded to a reference grid $\Omega$ (all FoV) 
and a single common system 
by applying a set of interlinking equations (\ref{eq:cond2}).
It can be shown that the final solution in $\Omega$ does not
depend on the size of the initial frames $\omega(i,R)$.
Residuals $V$ of this solution in each
image $m$ meet conditions  $ \vec {V^{\rm T} {\bar P} f = 0}$.
These conditions correspond to Eq.(\ref{eq:ort}) for reference stars
in $\Omega$ providing that  $P_{im}={\bar P}_i$.
Therefore a solution with
overlapping  reference subframes is equivalent to that
in a single isolated area  $\omega(i,\infty)=\Omega$,
or to a standard solution performed with all stars used as reference only.
Consequently,
no improvement in the  signal to noise ratio is expected.
This version of the reduction is useful for 
a low number of model parameters (vector $ \vec {c}$ is not used
in the model), high uniformity of model parameters and residuals $V$,
and a fast convergence of iterations. However, assumption
$P_{im}={\bar P}_i$ used here means that $P_{im}$ is constant in time,
which is not always acceptable.

For this study,  we used standard reduction
(Sect.3.3) computing the  target position with weights 
$P_{0}=0$, ${\bar P}_0=0$  to ensure the best  response 
to systematic errors in $V$.

%\newpage

%%%%%%%%%%%%%%%%%%%%%%%%%%%%%%

\section{Astrometric data reduction}

\subsection{Reduction of FORS1 images}

One of the FORS1 sky images
obtained in  Dec 2000 at a seeing near its mean level was used as a reference.
Photocenters were computed for 180 stars of $B=$18--24 mag 
in the central area.
Reductions were performed with a standard model (Sect.3.3)  that 
yields residuals $V$ and model parameters $\xi$ 
of a target star $i$ relative to
the local grid of reference stars $\omega(i,R)$. 
Due to  the extremely small value of systematic errors, we tried
to improve the statistics by accumulating data over all 
stars available. 
Therefore, reductions were repeated 180 times, processing  each star $i$
in turn, as a target (which till now was denoted by a subscript $i=0$).
For that reason, the astrometric precision varied depending on 
the distance $r$ of the star $i$ from the frame center.
For the brightest stars, this occurred due to
$\gamma$ decreasing from 
$\gamma \approx 0.6$ at the frame center to 0.2 at the periphery
(Fig.\ref{gam}) with a corresponding error increase by 25--50\%. 
Computations were  carried out with all $k$ 
from 6 to 16 and all $R$ from some $R_{\rm min}$ (Table \ref{RR}) that
provides  the minimum number of
reference stars needed at a given $k$, to the maximum $R_{\rm max}=2.3$\arcmin.
The first run was performed with zero image motion and afterwards 
computations were repeated with the actual estimate of $\delta$ (Sect.5).

\begin{table}[tbh]
\caption [] {Minimum and optimal radii of reference fields}
\begin{tabular}{rrr}
\hline
\hline
$k$     & $R_{\rm min}$  & $R_{\rm opt}$   \rule{0pt}{11pt}\\ 
\hline
6  &  40\arcsec   & 50\arcsec      \rule{0pt}{11pt}\\
8  &  50\arcsec   & 70\arcsec     \\
10 &  60\arcsec   & 90\arcsec     \\
12 &  70\arcsec   & 110\arcsec     \\
14 &  80\arcsec   & 140\arcsec     \\
16 &  90\arcsec   & $\sim$140\arcsec      \\

\hline
\end{tabular}
\label{RR}
\end{table}

We assumed a 10-parameter model for $\xi$ with zero-points, proper motions
$\mu^x$, $\mu^y$, atmospheric differential chromatic parameter $\rho$, 
the LADC compensating displacement $\rho'$, and with no parallax
for integer year differences between epochs.
Four extra $g_{xx}$, $g_{xy}$, $g_{yx}$, and $g_{yy}$ model terms
for each star were applied to compensate 
for a strong, over $\pm$200~px jittering of images.
Jittering induced a large signal in
$V$, clearly correlated with telescope
displacement $\Delta_ x$, $\Delta_ y $  along the $x$, $y$  axes.
This effect increased with $R$,
reaching several milliarcseconds at $R \approx R_{\rm max}$, $k \leq 8$.
The jittering makes the reduction
difficult, and is the reason that we discarded
a linear ($k=4$) model reduction.

Jittering moves the star field with reference to the camera optics.
This causes a change in the optical distortion (say along $x$)
at some point $x,y$ from its initial value ${\Phi}_x (x,y)$ to 
$\Phi_x(x,y)+ (\partial \Phi_x(x,y) / \partial x ) \Delta_ x + 
(\partial \Phi_x(x,y) / \partial y)  \Delta_ y$. A similar  expression
is valid for the distortion  $\Phi_y(x,y)$ along the $y$ axis.  Naming
the partial derivatives used here as $g_{**}$, we come to the expression
\begin{equation}
\begin{array}{ll}
\label{eq:optab}
{\Phi}'_x(x,y)= {\Phi}_x(x,y)+  g_{xx}  \Delta_ x + g_{xy}  \Delta_ y \\
{\Phi}'_y(x,y)= {\Phi}_y(x,y)+  g_{yx}  \Delta_ x + g_{yy}  \Delta_ y, \\
\end{array}
\end{equation}
equally applicable for the correction of positions.  Assuming that
optical distortions are stable over the observing
period, $g_{**}$ terms are included in, and found as components of $\xi$.

\begin{figure}[htb]
\resizebox{\hsize}{!}{\includegraphics*[bb = 59 53 301 168]{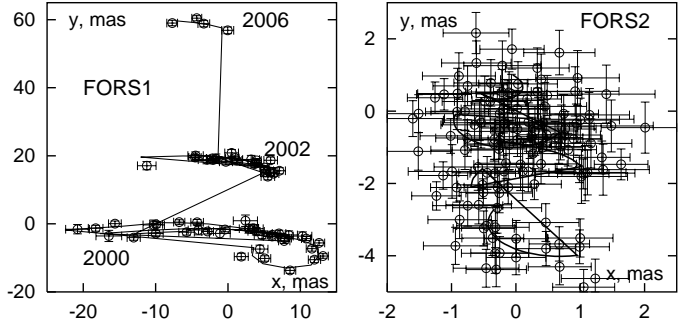}}
\caption {Example of $B=20.8$ star motion over a CCD traced for 
6 years with FORS1 and 5 months with FORS2:   measured
positions (open circles with error bars) and 
model track (solid curves). Reduction was performed
with parameters $k=10$ and $R=1.5$\arcmin.}
\label{motion}
\end{figure}
%%%%%%%%%%%%%%%%%%%%%%%%%%%%%%

An example of the measured and model star
motion over the CCD surface is shown in Fig.\ref{motion} 
for a red ($B=20.8$, $B-R=2.8$)  star with
proper motion $\mu^x=-1.32 \pm 0.06$~mas yr$^{-1}$, 
$\mu^y=10.74 \pm 0.06$~mas  yr$^{-1}$,
and trigonometric parallax 
(all relative) $\pi=0.286 \pm 0.053$~mas which was computed 
later based on  FORS2 images (Sect.4.2). This 
graph is actually rather simplified because it refers to positions 
corrected for polynomial $\vec{fc}$ and for jitter related terms.
The intricate shape of the track is due mainly to the DCR shift 
of images within  a single series. In the blue band, this motion exceeds
$\pm 10$~mas while in the red filter the effect is an order lower.
This makes it clear, for instance, why reference star displacements should be
taken into account when processing $B$ images. 

\begin{figure}[htb]
\resizebox{\hsize}{!}{\includegraphics*[bb = 52 49 267 172]{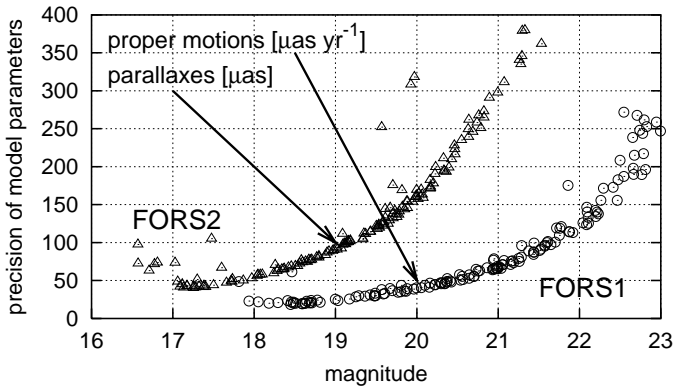}}
\caption { Astrometric precision of 
relative proper motions (open circles)
determined from FORS1 images with a six year time span 
and trigonometric parallaxes (triangles) derived from FORS2 images with
corrections based on  FORS1 proper motions. 
Reductions performed with $k=10$ and $R=1.5$\arcmin.}
\label{mupi}
\end{figure}
%%%%%%%%%%%%%%%%%%%%%%%%%%%%%%

The dependence on magnitude of the internal precision of proper motions  
is shown in Fig.\ref{mupi}. These estimates correspond 
to the formal least squares precision and take into account 
components  $\varepsilon$, $\delta$, and $\Delta_{\rm rf}$
of the total random error $\sigma$.
Due to the  large time span between epochs, proper motions were
derived with high precision, reaching $20$~$\mu$as yr$^{-1}$ for bright stars.
Systematic errors (Sect.6) degrade precision little 
since their contribution is small in comparison to random errors
(see Sect.4.2).

\subsection{Reduction of FORS2 images based on FORS1 proper motions}
For processing of FORS2 images we used high-precision proper motions
$\mu_{i}$ derived from the reduction of FORS1 images on a six year time base. 
At five-month spacing, they ensure very accurate corrections.
The possibility to use proper motions that fit  
image measurements of the other camera, different wavelength, and    
at distant epochs, however, is not evident and should be used with care. 
On the other hand, the precision of FORS2 positions obtained in this way
is an indicator of the actual accuracy of FORS1 proper motions.
Most important, the elimination of proper motions from 
model parameters
$\xi$  essentially increases the DoF (degrees of freedom)
of least square residuals in
temporal subspace thus making further 
study of systematic errors more reliable.

The reduction was started by finding the FORS2 image 
obtained at normal seeing
and best matching the star content of FORS1 images.
This image at epoch  $T_0$
was used as a reference for the reduction of all FORS2 images.
In most cases, the difference in star content 
from the two cameras occured for a small gap
between two CCD chips of FORS2 and
saturation of bright stars in the $R$ filter. 
In this way,  152 common stars were selected
for further reduction, which started from applying
corrections 
for proper motion occuring in star positions between epochs $T_0$
and $T_m$. 
This was performed taking into account
the singularity of astrometric reduction according to which
the $i$-th target position (and proper motion $\mu_{i}$) is related  
to a particular subset $\omega(i,R)$  of reference 
stars $j \in \omega(i,R)$, whose unique model parameters 
(denoted by $\mu'_{ij}$ in contrast to $\mu_{i}$)
are valid within this subset only. Therefore, 
to conserve the reference system, a reduction
of FORS2 images for a target $i$  was  performed 
with reference to the same frame $\omega(i,R)$ as used for FORS1. 
Also, we applied 
weights  ${\bar P}_i$ that are same as those involved in the reduction of
FORS1 images, that is, using average light fluxes in the blue filter.
Thus,  each reference (with respect to target $i$) 
star $j \in \omega(i,R)$ positions were corrected by
\begin{equation}
\label{eq:mured}
\Delta_{jm}  =- \mu'_{ij} (T_m-T_0).
\end{equation}
This complicated procedure is due to the necessity to conserve
the system of model parameters when processing different sets of images.
Violation of this principle immediately destroys the accuracy.
Thus, direct application of corrections $\Delta_{jm}=- \mu_{j} (T_m-T_0)$
to all measurements of reference stars in $\omega(i,R)$ 
is incorrect because these $\mu_{j}$ are related to their own
frames 
$\omega(j,R)$ which differ from $\omega(i,R)$. Mismatch of these areas 
and even a small inconsistency
of the reference star ensemble  sometimes result in large  
1000--5000~$\mu$as epoch residuals. Even use of proper motions
$\mu'_{ij}$ did not ensure complete identity of
$\omega(i,R)$ related to each camera due to inavailability of some FORS1 stars
in FORS2 images. 

The reduction model included zero-points, chromatic parameters $\rho$, $\rho '$,
and trigonometric 
parallaxes $\pi$. Formal random precision of FORS2 parallaxes for stars
of different brightness is given by Fig.\ref{mupi}.
For the best  stars, 
relative parallaxes are
determined with a precision near to  40~$\mu$as, 
which means that distances at 1 kpc are measurable with
a 4\% precision. A few large upward deviations in Fig.\ref{mupi} 
for some stars are caused
by a low number of measurements (oversaturation of bright images,
or position in the gap between two chips of the camera)
or by the peripheral position of stars and thus low $\gamma$ 
(large $\Delta_{\rm rf}$).

The precision of parallaxes is increased by use of FORS1 proper
motions, allowing us to exclude a component $\mu^x$  from model
parameters $\xi$, 
removing in this way a strong correlation between $\pi$ and $\mu^x$. In other
cases, the expected 
precision of parallaxes from the 5-month series of observations is
$200$~$\mu$as only.

Systematic errors, of course,  affect the accuracy of both proper motion
and parallax determination. In Sect.6 we show that the systematic
error for targets near frame center is about $25$ $\mu$as, or
a half of the  random error of epoch average positions for bright stars,
at months to year time scales.
Translating this estimate to parallaxes, we find that
 systematic errors contribute approximately $\pm20$~$\mu$as to each star parallax
and $\pm10$~$\mu$as yr$^{-1}$ to proper motions irrespective of the star
magnitude.

%\newpage
%%%%%%%%%%%%%%%%%%%%%%%%%%%%%%

\subsection{Merged residuals $\langle V\rangle $}

\begin{figure}[htb]
%   \centering
\begin{tabular}{@{}c@{}}
\resizebox{\hsize}{!}{\includegraphics*[]{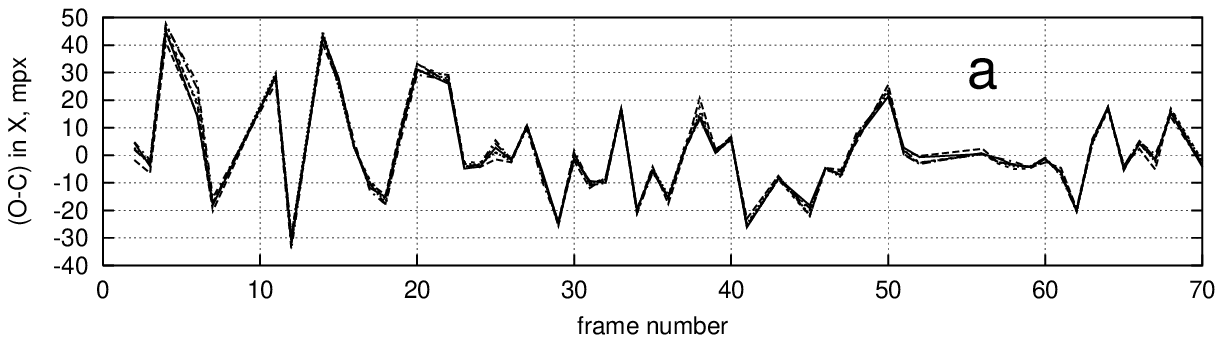}}\\
\resizebox{\hsize}{!}{\includegraphics*[]{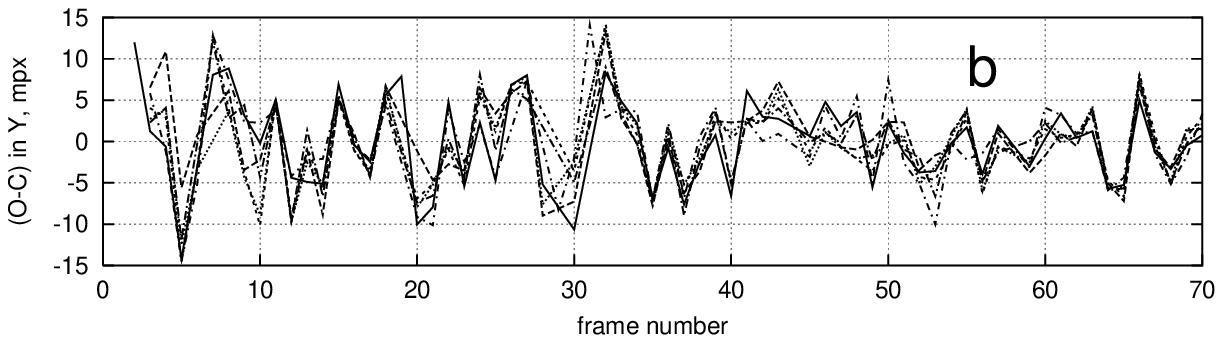}} \\
\end{tabular}
\caption {Image-to-image change of  FORS1 residuals $V$
for   {\bf a)} a faint 22 mag star ($\varepsilon_{}= 20 {\rm mpx}=2000\mu$as)
and {\bf b)} bright 19 mag star ($\varepsilon_{}= 3  {\rm mpx}=300\mu$as);
various line types refer to reduction parameters $k$ from 6 to 16. Computations
were made at  $R_{\rm opt} $.
}
\label{oc}
\end{figure}
%%%%%%%%%%%%%%%%%%%%%%%%%%%%%%

While processing,  we computed
residuals $V$ for each star $i$, each image $m$,
all reduction modes $k$ from 6 to 16, and several 
reference field sizes $R$, including $R_{\rm {opt}}$. 
The best precision 
residuals computed at  $R=R_{\rm {opt}}$ we denote as $V_{im}(k)$.
In practice, however, 
we do not require multiply defined residuals but rather a {\it single }
set of residuals 
which for a particular star $i$ is the best estimate of the planetary 
signal at the moment of image $m$ exposure. For that purpose we merge
$V_{im}(k)$ into a single system, the possibility of which follows from
the discussion in Sect.3.5.

\begin{figure}[htb]
\resizebox{\hsize}{!}{\includegraphics*[bb = 52 49 303 171]{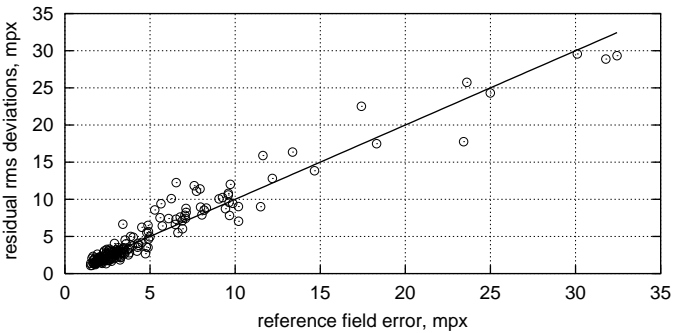}}
\caption { Standard  deviation (scatter with respect to $k$) of
$V(k)-\langle V\rangle $ for each FORS1 star as a function
of $\Delta_{\rm rf}$ (open circles) and a linear approximation (solid line).
}
\label{coroc}
\end{figure}

Let us consider
Fig.\ref{oc} that presents an example of the image-to-image
change of  $V_{im}(k)$ computed with  different $k$ 
for two stars of different brightness.
Residuals corresponding to different $k$ are seen to be highly correlated,
especially for a faint star,
and fluctuate near their average,  $\langle V\rangle_{im} $
being a function of $m$.
Recall that according to (\ref{eq:systV}), 
$  V_{im}(k) = {V}_{{\rm abs},{im}}- \sum'_{j} a_{ij}(k) {V}_{{\rm abs},{jm}}$
where $a_{ij}(k)$ refer to $k$ used.
Therefore 
$\langle V\rangle _{im} = {V}_{{\rm abs},{im}}- 
\sum'_j \langle a_{ij} \rangle {V}_{{\rm abs},{jm}}$ where 
$\langle a_{ij} \rangle $ is an average of $a_{ij}(k) $ with respect to $k$.
Hence $  V_{im}(k) - \langle V\rangle _{im}= 
\sum'_{j} [a_{ij}(k)-  \langle a_{ij} \rangle ] {V}_{{\rm abs},{jm}}$.
The variance of this difference, neglecting the second term, is
$\sum'_{j} [a_{ij}(k)]^2 \tilde{\sigma}^2_{jm}$, or 
$\Delta_{\rm rf}^2(k) $ at $k$ given.
Thus, the standard deviation of 
$V_{im}(k) - \langle V\rangle_{im} $ depends on  $\Delta_{\rm rf}$
almost linearly. This approximation is confirmed by actual data,
as shown in Fig.\ref{coroc}.

%%%%%%%%%%%%%%%%%%%%%%%%%%%%%%

Given 6 sets of ${V_{im}}(k)$ corresponding to  $k=6...16$ for each target $i$,
we merged them into the 
weighted average $\langle V\rangle_{im} $ using
weights $\Delta_{\rm rf}^{-2}$. Along with $V$, 
merged residuals $\langle V\rangle $ were
tested  for the presence of systematic errors (Sect.6).
As explained in Sect.3.5, the merging is not applicable to model
parameters $\xi$.

For faint stars, the relative amplitude of  $V$ fluctuations near 
$\langle V\rangle $ is insignificant  (Fig.\ref{oc}a) since  
$\Delta_{\rm rf} \ll \varepsilon_{}$. Therefore  
$\langle V\rangle \approx V$ at any $k$ and the use of  $\langle V\rangle $ 
instead of $V$ is of low efficiency. For bright stars (Fig.\ref{oc}b),
the precision of $\langle V\rangle$ is better 
due to the  averaging of the reference frame noise.

%\newpage

%%%%%%%%%
\section{Random errors}

\subsection{Calibration of the image centroiding error
          $\varepsilon_{}$ dependence on flux}

In this Section, our study was carried out with
images obtained in a narrow seeing range of 0.47--0.78\arcsec, which
includes almost all FORS1 and about 80\% of FORS2 images. The use of 
images out of this range leads to 
a noticeable increase of random errors.

\begin{figure}[htb]
\begin{tabular}{@{}c@{}}
\resizebox{\hsize}{!}{\includegraphics*[bb = 51 61 282 159]{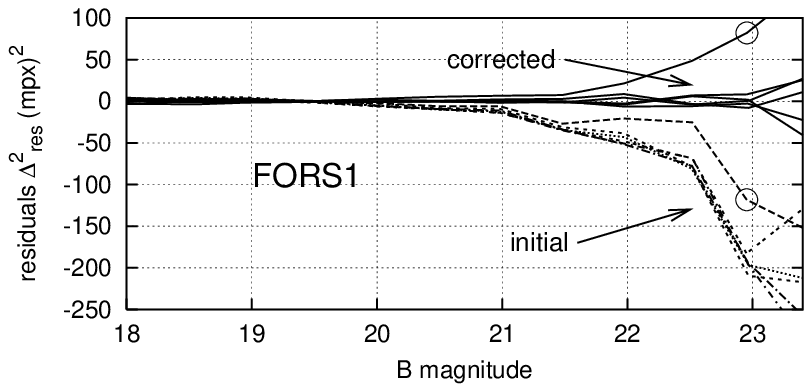}}\\
\resizebox{\hsize}{!}{\includegraphics*[bb = 51 48 282 159]{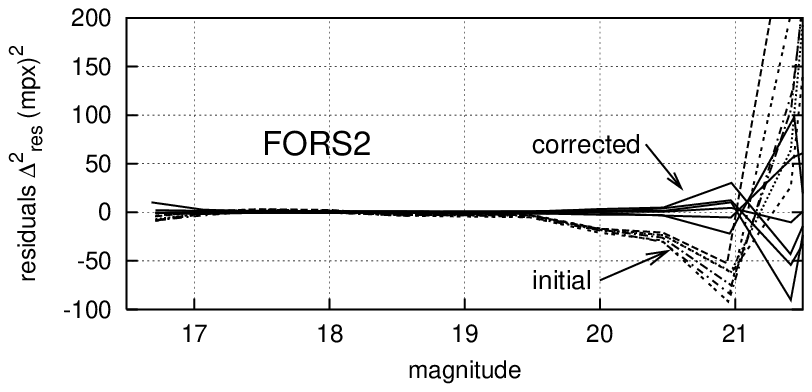}} \\
\end{tabular}
\caption { Initial (dashed lines) and corrected (solid lines) residuals 
of the error expansion (\ref{eq:res0}) 
computed at minimum reference field size $R_{\rm {min}}$. 
Different lines correspond to reduction with $k=6..\ldots 16$.
The case of $k=6$  is marked by open circles  
(deviating pair of lines for FORS1).}
\label{modres}
\end{figure}
%%%%%%%%%%%%%%%%%%%%%%%%%%%%%%

\begin{figure}[htb]
%   \centering
\begin{tabular}{@{}c@{}}
\resizebox{\hsize}{!}{\includegraphics*[]{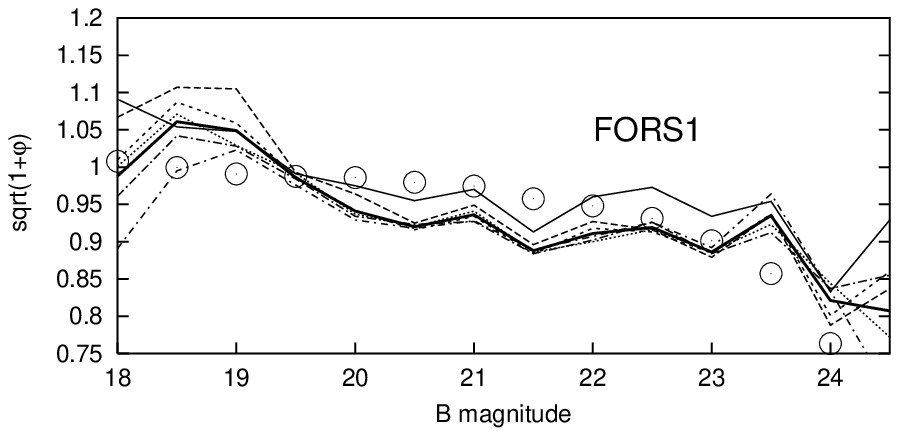}}\\
\resizebox{\hsize}{!}{\includegraphics*[]{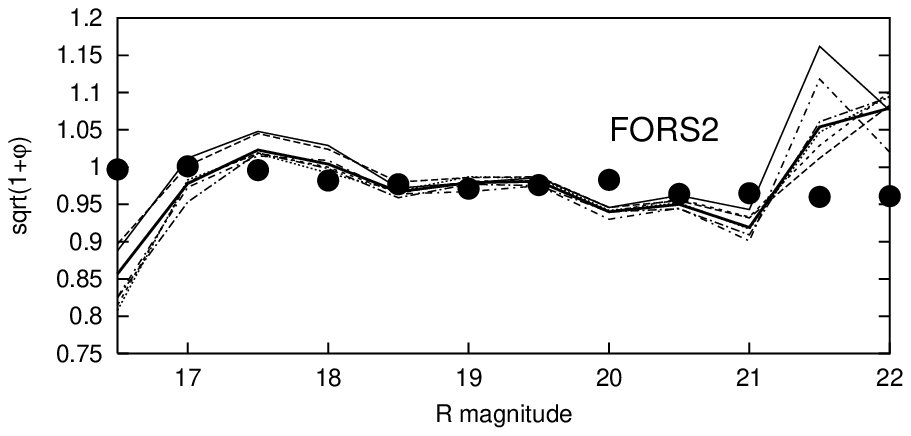}} \\
\end{tabular}
\caption { Correction $\sqrt{1+\varphi}$ for the model error 
of photocenter measurements $\varepsilon$. Estimates
for each $k$ reduction parameter (dashed lines) and their average 
(thick solid lines). Opened and filled circles reproduce
the ratio of image size (Fig.\ref{soni}) in the selected and initial star
sample}
\label{exsigma}
\end{figure}

The use of stars of different brightness to
investigate systematic errors requires careful
calibration of 
the dependence on flux of the image centroiding error $\varepsilon_{}$ 
(\ref{eq:phsigma}).
For calibration purposes, the best 
residuals are $V$ computed at the minimum possible $R=R_{\rm {min}}$ (Table \ref{RR})
since they
contain negligible input of atmospheric image motion $\delta =0$.
Using the variance $\sigma_{i}^2$ of residuals $V$ computed
for each $k$ at  $R=R_{\rm {min}}$, 
we can find the residual discrepancy of the decomposition (\ref{eq:vv}) 
into error components 
\begin{equation}
\label{eq:res0}
\Delta_{\rm{res}}^2=\sigma_{i}^2- \varepsilon_{i}^2-\Delta_{\rm rf}^2+
   [\vec {\nu^{\rm T} N^{-1}(i) \nu } ]_{mm}
\end{equation}
for each target $i$. The change of this quantity with star magnitude is shown 
in Fig.{\ref{modres}}  by dashed lines for each $k$.  
All curves corresponding to different $k$ modes closely follow 
a common dependence with  little scatter. 
The anomalously large deviation seen for the  FORS1 camera  at $k=6$ originates from
the  large jittering of images which  
was not completely compensated by the reduction. For high $k$ modes 
this effect is well removed.
While for bright stars, discrepancies $\Delta_{\rm{res}}^2$ are fairly small, 
at the faint end we note a systematic negative bias caused by
incorrect modelling of errors. 
This bias almost does not depend on
$k$ and is approximately proportional to
$\varepsilon_{}^2$. Therefore we assumed that
this discrepancy is caused by insufficient validity of model 
(\ref{eq:phsigma}) for $\varepsilon_{}^2$, which requires
an additive correction $\varphi \varepsilon_{}^2$ 
with a coefficient $\varphi$ independent of flux. 
A similar correction should be applied to
$\Delta_{\rm rf}^2$ also. Correction factors $\sqrt{1+\varphi}$
to $\varepsilon_{}$ and $\Delta_{\rm rf}$  computed 
in 0.5~mag flux bins are shown in Fig.{\ref{exsigma}} as a function
of magnitude.

A change of  $\sqrt{1+\varphi}$ with brightness in Fig.{\ref{exsigma}}
is similar for both cameras.  A negative trend 
over a 4--5~mag range of brightness  
reproduces the dependence of star image size 
$\sigma_{\rm{G}}$ on flux (Fig.\ref{soni}) and therefore probably
is a consequence of selective filtration
based on star profile parameters when star images with excessive size
were discarded (Sect.2). 
The use of more compact images in comparison to 
the initial star sample, of course, results in an
improvement of the effective centroiding error $\varepsilon_{}$ 
observed in Fig.{\ref{exsigma}}. 
A similar improvement of  precision for the brightest images
occurs for the selection based on $\chi^2$  criterion (Sect.2).

Averaging with respect to $k$ produced final estimates 
$\sqrt{1+\varphi}$ shown in Fig.{\ref{exsigma}} by solid lines. 
With these corrections, residuals 
(\ref{eq:res0}) have been recomputed yielding new 
discrepancies  $\Delta_{\rm{res}}^2$ with much smaller 
magnitudes (Fig.{\ref{modres}}, solid lines). 
Having found the calibration factor $\sqrt{1+\varphi}$, we can
correctly estimate $\sigma_{ {V}}$  at any
$R>R_{\rm min}$ simply by adding the  image motion variance
$\delta^2$: 
\begin{equation}
\label{eq:modV2}
\sigma_{ {V}}^2= 
  (1+\varphi)\varepsilon_{i}^2+ (1+\varphi)\Delta_{\rm rf}^2 + 
    \delta^2_i +\Delta^2_{\rm{res}} -
    [\vec {\nu^{\rm T} N^{-1}(i) \nu } ]_{mm} .
\end{equation}
The term  $\Delta_{\rm{res}}^2$ is used to take into account
the dependence of $\varphi$ on  flux which originally was
considered constant. This also compensates, at least
statistically, the use of a single correction factor 
for both $\varepsilon^2$ and $\Delta_{\rm rf}^2$.

\begin{figure}[h]
\resizebox{\hsize}{!}{\includegraphics*[]{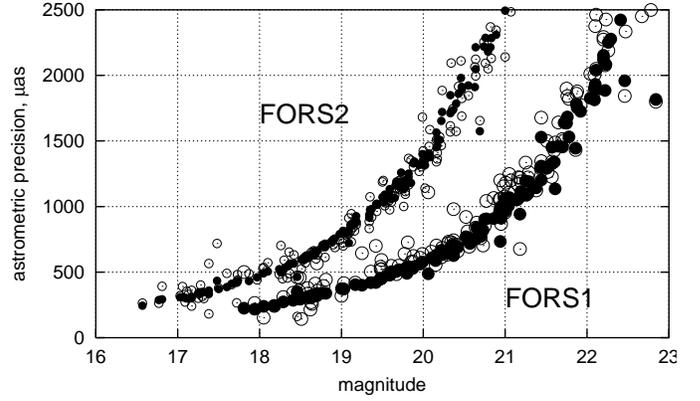}}
\caption { 
Astrometric precision of a single photocenter measurement: 
observed (open circles) and model estimate
$\varepsilon\sqrt{1+\varphi}$ (filled circles) for 
FORS1 (large symbols) and FORS2 (small symbols) as a function
of magnitude. 
}
\label{sig12}
\end{figure}

The validity of above calibration
is illustrated in Fig.\ref{sig12} where we compare
the astrometric precision of a single photocenter measurement 
restored from  observations with its model prediction
$\varepsilon\sqrt{1+\varphi}$ in the case of 
reductions with $R=1.5$\arcmin and $k=10$. 
The  measured astrometric precision, for each star, was computed 
based on the observed variance $\sigma_{ {V}}^2$
of $V$ (mean in $x$ and $y$),
$ \delta^2 $ derived in Sect.5.2, and representation (\ref{eq:modV2}).
These results,  as for model values $\varepsilon\sqrt{1+\varphi}$
for each target,
were averaged over all data available. 
Fig.\ref{sig12} shows a good match of the observed
and model precision over wide range of magnitudes. 
This graph matches well  our previous results for FORS1 based
on a reduction technique with overlapping reference frames 
(Lazorenko et al. \cite{Lazorenko7}).

We emphasize  that both $\varepsilon\sqrt{1+\varphi}$ and 
$\varepsilon$  are estimates of the 
actual precision of the photocenter determination.  The difference is
that the first one refers to
the star sample affected by {\it selection } while $\varepsilon$  is 
related to 
the imaginary sample of FORS images with no defects.
In spite of the  small value of $\varphi$, the subsequent study of
image motion and  systematic errors greatly favours its use since it allows us
to incorporate large amount of data from faint stars.

%%%%%%%%%%%%%%%%%%%%%%%%%%%%%%

\subsection{Image motion}

\begin{figure}[ht]
\begin{tabular}{@{}c@{}}
\resizebox{\hsize}{!}{\includegraphics*[]{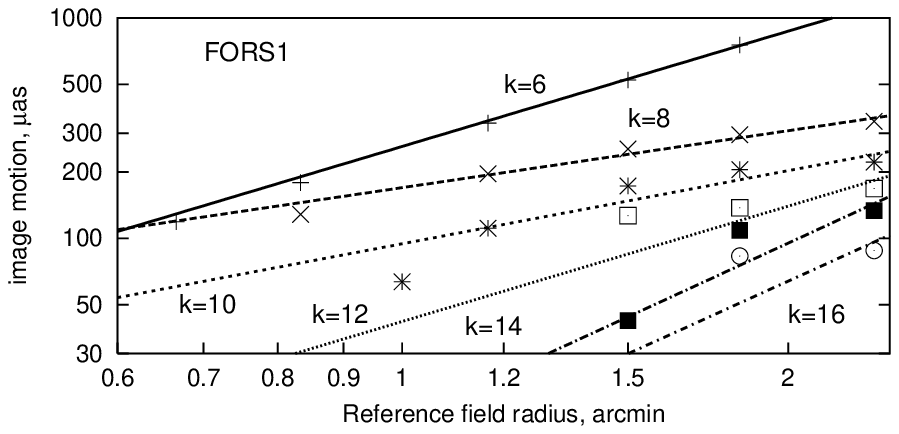}}\\
\resizebox{\hsize}{!}{\includegraphics*[]{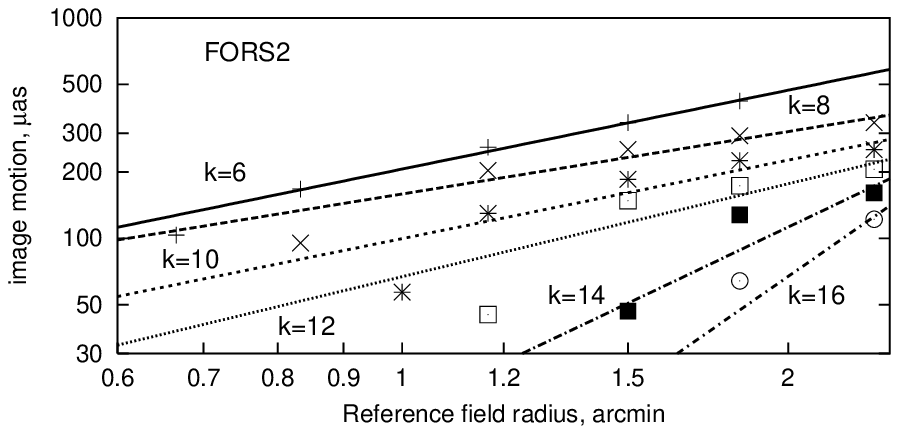}} \\
\end{tabular}
\caption { Image motion $\delta$ as a function of reference field size $R$ 
computed with $k=6 \ldots 16$ (symbols of different type) and 
corresponding fits (lines) by a power law (\ref{eq:atmr}).  }
\label{atm}
\end{figure}
%%%%%%%%%%%%%%%%%%%%%%%%%%%%%%

Taking advantage of the  availability of a well calibrated image centroiding error,  
we used  Eq.(\ref{eq:modV2}) to extract the image motion  
component $\delta$ at various $R$.     
This equation was solved numerically for each star
taking into consideration the fact that the reference frame noise
$\Delta_{\rm rf}$ is a function of  $\varepsilon$ and $\delta$.
The results averaged over all stars available  and computed for each $k$
and $R$ are  shown in Fig.\ref{atm}.
Comparing estimates 
obtained for both  cameras, one may note the  similarity of results
in spite of the 
difference in pixel scale, number and spacing of epochs, 
different reduction model parameters, different method of reduction and, 
especially, an 8-fold difference in
exposure $T$ (600 and 70 sec for FORS1 and FORS2 respectively). 
The last aspect raises a doubt about the validity 
of relating the measured image motion to atmospheric turbulence.

At each  fixed $k$,  $\delta$ estimates were fitted by a power law
\begin{equation}
\label{eq:atmr}
        \delta=B R^{b} 
\end{equation}
assuming that $R$ is given in minutes of arc.
Fitting parameters $B$ and $b$ 
are given in Table \ref{atmtable} for the first few $k$ modes only
since the results for $k>12 $ are too uncertain. 
Excessive estimates of $B$  (in comparison to FORS2) found at 
$k=6$ and $k=8$ could be due to the residual
effect of large image jittering of FORS1 images.
For comparison,
the table contains  $B_{\rm a}$ and $b_{\rm a}$ coefficients of 
Eq.(\ref{eq:atmr}) expected for differential image motion
caused by atmospheric turbulence. These values were obtained by scaling
model coefficients (Lazorenko \& Lazorenko \cite{Lazorenko4}) 
that refer to
typical atmospheric conditions at Chilean observatories,
to current exposures and telescope aperture.
The atmospheric-related amplitude $B_{\rm a}$ is much smaller than the observed one,
especially for FORS1 with a 3-8 fold discrepancy.  
Such a large difference suggests that we are measuring an effect not related to 
the atmospheric turbulence.

In a pilot study of FORS2 astrometric performance, Lazorenko (\cite{Lazorenko6}) 
estimated  Eq.(\ref{eq:atmr}) 
parameters  $B'$ and $b'$
using a single night observation series with $T=17$ s exposure. 
Coefficients $B'$ reproduced in Table \ref{atmtable}  are
approximately twice as large as in  this study, possibly 
due to the different technique of reductions, which now
takes into
account DCR displacement of reference stars.

\begin{table}[tbh]
\caption [] {Coefficients  of Eq.(\ref{eq:atmr}):
derived in this study $B$[$\mu$as], $b$; 
predicted $B_{\rm a}$, $b_{\rm a}$ 
by atmospheric model (Lazorenko \& Lazorenko \cite{Lazorenko4});
and $B'$, $b'$ obtained from 
a single series of FORS2 images (Lazorenko \cite{Lazorenko6}) }
\begin{tabular}{@{}r|rr|rr|rr|rr|rl@{}}
\hline
\hline
  &\multicolumn{4}{|c|}{ FORS1 } & \multicolumn{6}{|c}{ FORS2 } \rule{0pt}{11pt}\\
\hline
$k$& $B$  & $b$ & $B_{\rm a}$ & $b_{\rm a}$ & $B$   & $b$  & $B_{\rm a}$ &   $b_{\rm a}$ & $B'$ & $b'$ \rule{0pt}{11pt}\\ 
\hline                 
6  & 261*  & 1.74& 31    & 1.6   &  206  & 1.19 & 92    &  1.6    & 380  & 1.2   \rule{0pt}{11pt} \\  
8  & 170*  & 0.86& 27    & 1.8   &  159  & 0.94 & 79    &  1.8    & 340  & 1.2  \\                    
10 &  94  & 1.10& 14    & 1.7   &  100  & 1.18 & 41    &  1.7    & 205  & 1.6  \\                    
12 &  42  & 1.74& 13    & 1.8   &   67  & 1.40 & 37    &  1.8    & 180 & 1.7 \\                     
\hline
$T$ &\multicolumn{2}{|c|}{600 s}&\multicolumn{2}{|c|}{ 600 s}&\multicolumn{2}{|c|}{70 s} &\multicolumn{2}{|c|}{70 s} &\multicolumn{2}{|c}{17 s}\rule{0pt}{11pt}\\
\hline
\end{tabular}
\begin{list}{}{}
\item*  Could be biased due to the residual
effect of large image jittering.
\end{list}
\label{atmtable}      
\end{table}

In all cases, the measured  powers $b$ of Eq.(\ref{eq:atmr}) are
significantly below their predicted values $b_{\rm a}$. 
We conclude that the observed image motion at $T \geq 70$ s is not due to 
atmospheric turbulence since it does not decrease as $T^{-1/2}$
and therefore is of {\it instrumental} origin. 
Very likely, it does not depend on exposure, at least 
for $T \geq 70$ s. Due to domination over 
the intrinsic atmospheric image motion, the last component 
not can be extracted from the present data.
In Sect.6 we suggest that the image motion detected probably
is caused by star image asymmetry in combination with variations of the PSF.

%%%%%%%%%%%%%%%%%%%%%%%%%%%

%\newpage

\section{Systematic errors at monthly/annual epochs}

In our former study (Lazorenko et al. \cite{Lazorenko7}),
systematic errors 
in positional observations with the FORS1 camera were shown  
to be about 30~$\mu$as. 
The detection of such weak signals presents a certain 
difficulty and limits our search to characterization
of the error component
invariable within each monthly/annual epoch of observations.
Systematic signs in observations may appear for effects
not described by the reduction model. The most troublsome are 
long-term instabilities
which differently affect images at distant epochs,
e.g. changes in VLT optical aberrations, star colours, actual PSF shape,
variable background  gradient due to light from nearby stars, etc.

\subsection{Epoch average residuals}
%%%%%%%%%%%%%%%%%%%%%%%%%%%%%
\begin{figure*}[ht]                                    
\centering
\includegraphics[bb = 59 53 532 157, width=17cm]{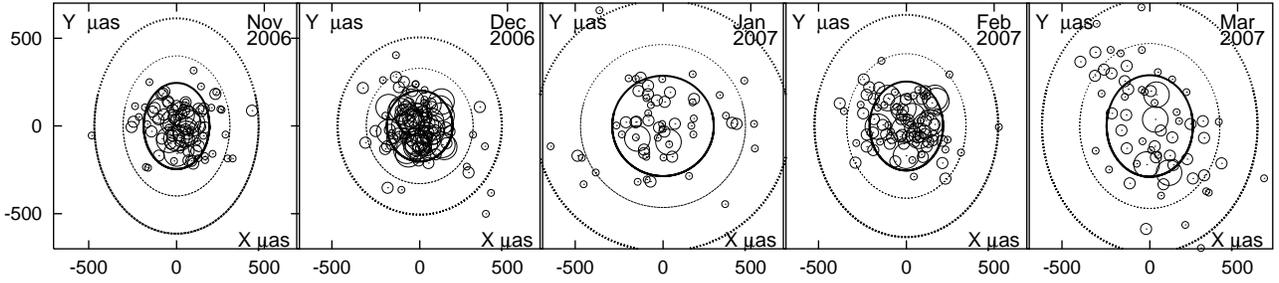}
\caption {Monthly 
normal points (average residuals) $V_e$ for stars
observed with FORS2. Symbol size refers to 
the expected  precision of normal points $D_e^{1/2}$,
which changes from better than 100~$\mu$as
(large circles, usually brightest stars), 
to 170~$\mu$as (middle), and  260~$\mu$as (small circles). 
Corresponding 3$\sigma$ scatter areas are shown by
dashed ellipses with semiaxes $3\theta_e D_e^{1/2}$ whose size
depend primarily on seeing (best in December 2006 and worst in January and
March of 2007). No large dispersions are seen. 
}
\label{dxdyAA}
\end{figure*}
%%%%%%%%%%%%%%%%%%%%%%%%%%%%%%

The quantities best suited to this study are the epoch
average weighted residuals
\begin{equation}
\label{eq:xe}
V_e=\sum \limits _{m \in e}V_{im} \sigma_{im}^{-2}  /
     \sum \limits _{m \in e}\sigma_{im}^{-2}         
\end{equation}
computed for each star 
at each monthly/annual observation epoch $e$.
A set of these epoch points $V_e$ was investigated to detect and characterize
systematic errors. 
The epoch normal points $V_e$ are formally 
characterized by variances 
\begin{equation}
\label{eq:De}
D_e= 1/\sum \limits _{m \in e}\sigma_{im}^{-2}
\end{equation}
equal to the cumulative weight of individual residuals of images
$m \in e$ available at epoch $e$.
However, variances $\hat{D_e}$ describing the  actual scatter of the normal points $V_e$
are below $D_e$ as a consequence of the  least squares fit.
Thus, assuming
a normal law for the  distribution of observation errors,
from Eq.(\ref{eq:xe}) we find  
\begin{equation}
\label{eq:Deoc1}
\hat{D_e}= \sum \limits _{m,m' \in e}B_{mm'}\sigma_{im}^{-2}\sigma_{im'}^{-2}
        /( \sum \limits _{m \in e} \sigma_{im}^{-2})^2
\end{equation}
where $B_{mm'}$ are diagonal elements of a covariance matrix 
\begin{equation}
\label{eq:Deoc}
         \vec{B}(i)=\vec{P^{-1}(i)} - \vec{\nu^{\rm T} N^{-1}(i) \nu}
\end{equation}
of residuals $V_{im}$ for $i$-th star and 
$\vec {P^{}(i)}$ is a diagonal matrix with elements
${\sigma}^{-2}_{im}$ introduced by Eq.(\ref{eq:var}).  
It follows that  $\hat{D_e}=D_e$ 
only when the second item in Eq.(\ref{eq:Deoc}) is zero. In practice,
instead of  Eq.(\ref{eq:Deoc1}), it is convenient to use the  expression
\begin{equation}
\label{eq:beta}
 \hat{D_e}=\theta_e^2 D_e  
\end{equation}
where $\theta_e \leq 1$ is a quantity numerically computed 
for a particular distribution of 
$i$-th star observations over time.
Although matrices $\vec { N^{-1}(i)}$ are unique  for each star, 
a minor difference in $\theta_e$ for 
different stars often can be neglected. 
Of course, we have two sets of $\theta_e$ 
values related to $x$ and $y$ axes. 

It is difficult to
suppose that systematic components follow exactly
the parallax and proper motion displacement of stars.
Therefore, after a fit in time, systematic errors add an extra scatter to
epoch residuals $V_e$, which is detected as an
excess in the expected value of the variance $\hat{D_e}$. 
This excess
we find below based on the well-calibrated  (Sect.5) model of 
the stochastic-dependent component of the variance.

A good idea of analyzed epoch average residuals $V_e$  is given
by graphs of Fig.\ref{dxdyAA}. This plot shows the 
typical distribution of 
monthly  normal points $V_e^{(x)}$, $V_e^{(y)}$ 
in the $x,y$ plane for each month, for stars
observed at FORS2, and the reduction with $k=10$ and
$R=R_{\rm {opt}}$. This distribution of $V_e$ is typical also
for processing with other  $k$ at $R_{\rm {opt}}$ due to 
the  high degree of correlation between these sets of residuals (Sect.4.3). 
Normal points are shown by open circles 
of three size grades which  refer to the
precision $D_e^{1/2}$ of normal points
better than 100~$\mu$as (largest circles, usually brightest central stars), 
170~$\mu$as  (middle size),  
and  260~$\mu$as (small circles, faint or peripheral
stars).  Ellipses (dashed curves) with semiaxes 
$3\theta_e D_e^{1/2}$ mark 
3$\sigma$ scatter limits expected for least square residuals.
The different scatter of $V_e$ 
for different epochs is caused primarily
by  seeing (Table \ref{obs}), which is best for 
the second (most compact location of dots) 
and worst for the third  and last 
epoch. For the same reason,
only a few points  with  $D_e^{1/2}<100$ $\mu$as precision (large signs)
are seen for the middle and the last epoch since only the
best stars are measured well at bad seeing. 
Most 
normal points of each precision grade  are inside of the
corresponding 3$\sigma$ limits with no wide dispersions.

\begin{figure}[htb]
%\resizebox{\hsize}{!}{\includegraphics*[bb = 53 49 311 420 ]{12026f16.eps}}
{\includegraphics*[bb = 53 49 311 420, width=8.7cm, height=11.2cm ]{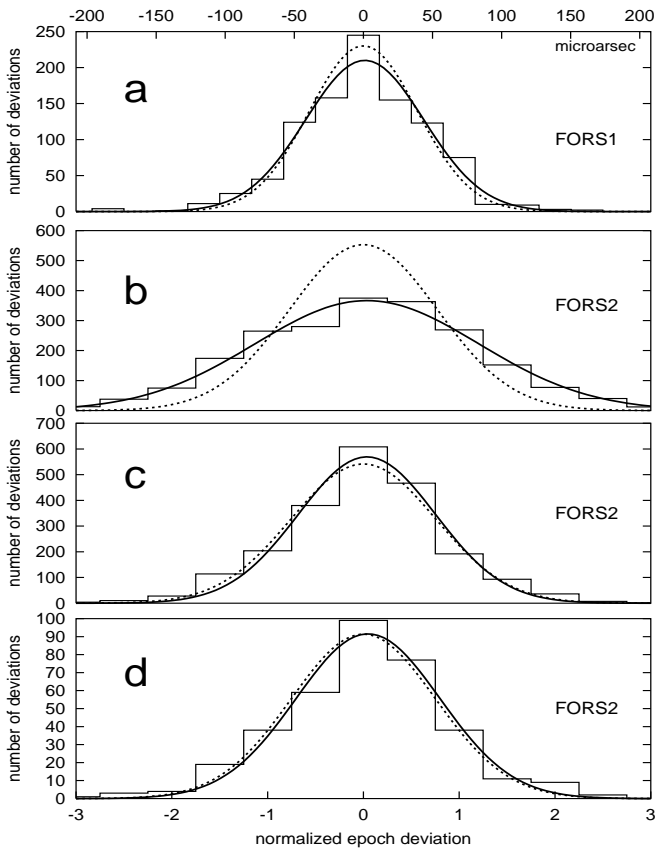}}
\caption { Histograms of $\overline{V_e}$ 
(epoch residuals ${V_e}$
 normalized to  unity to exclude dependence
on star brightness) in $r= 1$\arcmin central zone (steps),
Gaussian approximation (solid curves),
and theoretical distribution 
in the case of zero  systematic errors (dashed Gaussians), for: {\bf a)} -
FORS1;   {\bf b)} -  FORS2 complete set of images, 
with seeing varying from 0.38\arcsec to 0.84\arcsec;  {\bf c)} - 
FORS2 subset of images with 0.47--0.78\arcsec seeing;  {\bf d)} - 
merged FORS2 residuals $\langle V\rangle _e $.
Upper $x$-axis scale refers to the expected distribution of non-normalized
epoch residuals $V_e$ for brightest stars of $B$=18 mag for FORS1 
or $R$=16.5 mag for FORS2.}
\label{bin}
\end{figure}

Next, we  considered the frequency distribution
of $V_e$. These data however are not uniform in precision 
for different light fluxes from stars.
In order to exclude a dependence of the precision on brightness,
we introduced the dimensionless normalized quantities 
\begin{equation}
\label{eq:normv}
\overline{V_e} = V_e D_{e}^{-1/2}.  
\end{equation}
The standard deviation of $\overline{V_e} $, according to (\ref{eq:beta}), 
is equal to $\theta_e$ and therefore does not depend on brightness.
The frequency distribution of
$\overline{V_e}$ was formed cumulating data of all
epochs, model versions with $k= 6 \ldots 16$ at $R=R_{\rm opt}$,
for both axes,
and using bright $B<21$ stars in the central frame area $r<1$\arcmin. 
Histograms obtained and their Gaussian approximations
are given in Fig.\ref{bin}a,b for FORS1 and FORS2 respectively. 
These histograms are compared with the 
theoretical distribution of $\overline{V_e}$ 
in the  case of  zero systematic error and  
taking advantage of the fact that 
$\overline{V_e}$ follows a Gaussian distribution
with the variance parameter $\langle \theta \rangle^2$ equal to 
$\theta_e^2$ averaged over the epochs.

In the  case of FORS1,
from  Eqs.(\ref{eq:De}--\ref{eq:Deoc}) we find 
that typical values of $\theta_e^2$ are 
0.17, 0.48, and 0.27 for  epochs $e=1,2,3$ respectively,
with small variations depending on observing conditions of 
the particular target.
Therefore $\langle \theta \rangle=
 \sqrt{\sum \theta_e /3}= 0.56$
is a $\sigma$-width parameter for a theoretical Gaussian distribution.
The observed distribution of $\overline{V_e}$
for FORS1   is slightly wider,
with a 0.61 $\sigma$-parameter and a few large residuals  (Fig.\ref{bin}a).

In the case of FORS2, 
$\theta_e^2$ varies for different epochs between 0.22 and 0.76,
with an average $\langle \theta \rangle=0.78$.
The observed distribution is much wider, with a 1.17
$\sigma$-width parameter and 
a significant widening of wings  (Fig.\ref{bin}b).  The observed
distribution of monthly normalized residuals clearly indicates
the  presence of large systematic errors which is discussed later on. 

%%%%%%%%%%%%%%%%%%%%%%%%%%%%%%%%%%
%%%%%%%%%%%%%%%%%%%%%%%%%%%%%%
    
The histograms in Fig.\ref{bin} are sensitive to 
systematic errors providing their  magnitude is
comparable to the precision of epoch normal positions, about
50--200~$\mu$as. These histograms however are to be considered
primarily as illustrative.
Numerical characterization of systematic error is found
under the assumption that its value $A_e$ for a given star $i$
is constant within
each monthly/annual observation series $e$.  In this case
all measured residuals $V_{im}$ in images $m \in e$ are systematically 
biased by a constant $A_e$. Therefore
$V_e({\rm measured})= V_e({\rm at \; zero \;  systematic \; errors}) + A_e$ and 
the expectation of the variance of measured normalized residuals is
${\overline{V_e}}^2=\theta_e^2 + A_e^2 \theta_e^2  D_{e}^{-1}$.
The second item describing the input of  systematic errors 
in ${\overline{V_e}}$ dominates for bright stars.
This component was computed for each star and
averaged to derive statistically reliable
$A^2= \langle A_e^2\rangle$.
Averaging was performed over not too faint stars, all epochs, and 
all parameters $k$ at $R=R_{\rm opt}$,
assuming that the mathematical expectation of
$A_e^2$ does not depend on epoch, star light flux, and axis.
This yielded the representative estimate 
\begin{equation}
\label{eq:Ae}
A^2= \sum \limits  _{i,e}
              ( 0.5 \overline{V_e^x}^2+0.5 \overline{V_e^y}^2-\theta_e^2 )
              /  \sum \limits  _{i,e} ( \theta_e^2 /D_{e}).
\end{equation}

For a complete set of FORS2 observations (any seeing conditions),
and images in the central $r<1$\arcmin area,
we obtained $A=160$~$\mu$as.  
This value  exceeds by much the 
standard deviation for epoch average residuals of
bright stars, which is typically about 50-70~$\mu$as,
and therefore such   systematic errors  strongly
affect the histogram's shape (Fig.\ref{bin}b). 
It was found that large $A$ values are associated with images of
abnormally small FWHM, in particular FORS2 images in Dec 2006
with exceptionally good seeing of 0.3--0.5\arcsec.
Elimination of images with seeing below 0.47\arcsec (3.7~px)
and the  subsequent 
rejection of bad images with FWHM$>0.78$\arcsec (6.2~px) 
significantly improved  $A$. 
Such filtration produced 
better solution with essentially more compact histograms
of epoch-normalized residuals (Fig.\ref{bin}c) 
fitted with a Gaussian of 
only 0.72 $\sigma$ width parameter.  This is insignificantly
smaller than the expected $\sigma$-width which, due to
a change in $\vec{ N^{-1}(i)}$ caused by the above filtering,  decreased 
from  0.78 to 0.75.
The described filtering of 20\%  of the FORS2 images,
was applied to a few FORS1 images whose seeing was almost always
within  the limits adopted.
At the end of this Section we discuss the probable relation between
image size and systematic errors.

The frequency distribution of epoch average residuals $\langle V\rangle _e $
based on merged residuals $\langle V\rangle $
(Fig.\ref{bin}d)
does not differ from that built for $V_e$ (Fig.\ref{bin}c).

The estimates of $A$  for stars in CCD
central circular areas of $r=40$\arcsec, 1\arcmin, and 1.5\arcmin  \,
are given in Table \ref{TA}. 
%These computations for  FORS2 were performed 
%with a rejection of two moderately bright stars with abnormally 
%large individual $A$ to 400--500~$\mu$as . 

\begin{table}[tbh]
\caption [] { Systematic component $A$~[$\mu$as] 
in residuals $V_e$  
and in $\langle V\rangle _e$ within central CCD zones of $r$ radii}
\begin{tabular}{r|r@{}lr@{}l|r@{}lr@{}l}
\hline
\hline
  &\multicolumn{4}{c}{ FORS1 } & \multicolumn{4}{c}{ FORS2 } \rule{0pt}{11pt}\\
\hline
$r$    & \multicolumn{2}{c}{  in $V_e$}  & \multicolumn{2}{c}{in $\langle V\rangle _e$} 
       & \multicolumn{2}{c}{ in $V_e$ }  & \multicolumn{2}{c}{in $\langle V\rangle _e$}    \rule{0pt}{11pt}\\ 
\hline
40\arcsec   & 53&  $\pm 21$  &  25& $\pm 19$   & 65&  $\pm 22$  & 64 & $\pm 19$   \rule{0pt}{11pt} \\
1\arcmin   & 61 &  $\pm 9$   & 69 & $\pm 10$   & 68&  $\pm 9$   &  68& $\pm 10$   \\
1.5\arcmin  & 95&  $\pm 10$  &  79& $\pm 12$   & 85&  $\pm 6$   &  89& $\pm 6$  \\
\hline
\end{tabular}
\label{TA}
\end{table}

According to  Table \ref{TA}, the  characteristics of the systematic error $A$
for both  cameras
are similar and show a slight increase
in the direction from the center to the periphery of a frame.
Thus, while  at the periphery $A$ is  near to 100~$\mu$as,
at $r<40$\arcsec it does not exceed 50--60~$\mu$as.
Due to statistical limitations
we not can estimate $A$ at the center
(where the target is usually placed),
but considering the tendency observed we predict it could be about 25~$\mu$as
at $r<10$\arcsec,
as expected from the following discussion.
The systematic components in  $\langle V\rangle _e$ 
and $V_e$ are approximately equal.

\begin{figure}[htb]
  \centering
\resizebox{\hsize}{!}{\includegraphics*[bb = 61 53 316 184]{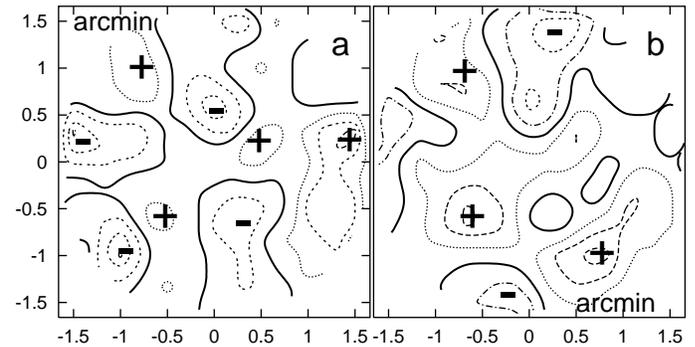}}
\caption { Global pattern of FORS1 systematic error distribution
over the CCD surface: {\bf a)} - in  normal points $V_e$  and  {\bf b)} -
in merged residuals $\langle V\rangle $ 
for a Dec 2002 epoch.
Isolines are plotted every 25~$\mu$as; zero level is shown by a solid line.
Residuals $V_e$ were computed with $k=10$ at optimal size of 
reference frames $R_{\rm {opt}}=1.5$\arcmin. }
\label{grid}
\end{figure}

\begin{figure}[htb]
  \centering
%\resizebox{\hsize}{!}{\includegraphics*[bb = 61 53 316 418]{12026f18.eps}}
{\includegraphics*[bb = 61 53 316 418, width=8.7cm,height=11.8cm]{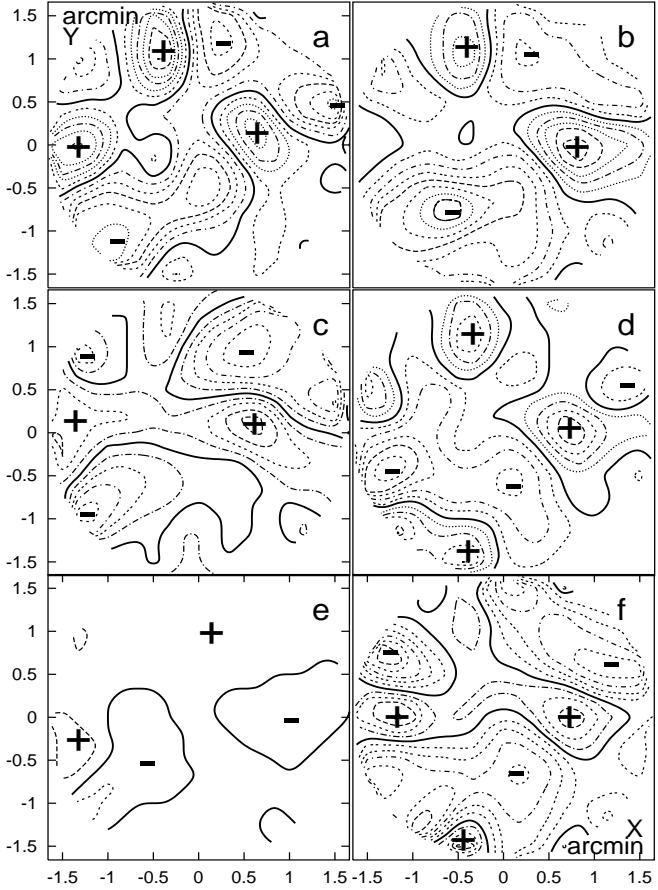}}
\caption {Distribution
over the CCD surface of a systematic component in FORS2 epoch residuals
in Jan 2007:
{\bf a)} -  in $V_e$ computed with $k=10$ and 
$R_{\rm {opt}}=1.5$\arcmin  with all 
images;  {\bf b)} - the same, for a subset of images in
a 0.47--0.78\arcsec seeing range;  {\bf c)} - the same, 
for a full set of images and modified reduction model
that takes into account dependence of positions on seeing  (Sect.6.2);  
{\bf d)} - for merged residuals $\langle V\rangle$;  {\bf e)} -
in $V_e$  at good seeing (Nov 2006);  {\bf f)} -
in $V_e$ (Jan 2007) for $k=6$  and $R_{\rm {opt}}=0.5$\arcmin.
Isolines are plotted every 25~$\mu$as; zero level is shown by a solid line.
See explanations in the text. }
\label{grid2}
\end{figure}

An important piece of information on the global distribution of 
systematic errors over
the  CCD  plane was derived applying a low-pass Gaussian spatial filter to
the epoch residuals ${V_e}$. 
The resulting low-frequency component $A(x,y)$ in  $V_e$ for each camera,
some epochs, and reduction versions
is shown in contour plots Fig.\ref{grid},\ref{grid2} with isolines
drawn with 25~$\mu$as increments. All graphs refer to the 
residuals on the $x$-axis.  
$A(x,y)$ function change is rather complicated and has
several extremums. This behavior, of course, not can  be
approximated by polynomials with basic functions $\vec{f}$
since this dependence is excluded in the course of the 
reduction procedure.   It is characteristic that 
large systematic errors reaching in most cases
100--200~$\mu$as  tend to concentrate at the periphery. 
At the field center, where the target is
usually placed,  $A(x,y)$ functions varies rather smoothly
and often fall to $<\pm 25$~$\mu$as. No graph was found to have
extremum at the center of the frame.

%%%%%%%%%%%%%%%%%%%%%%%%%%%%%%

In the case of FORS1, systematic error plots are given
for  middle 2002 epoch at which the maximum fluctuations of $A(x,y)$  
are detected. Fig.\ref{grid}a, plotted for $V_e$ computed with
$k=10$ at $R=R_{\rm {opt}}$ and Fig.\ref{grid}b for 
merged $\langle V \rangle$ residuals have few similar structures
above $\pm 50$~$\mu$as. At the center, errors are
negligibly small. Although we discuss here only
a particular case of the reduction with $k=10$ or 
of merged $\langle V\rangle $, our comments (both for FORS1 and FORS2) 
are valid also for computations with  other 
$k$ due to the  high correlation of results obtained
at $R=R_{\rm {opt}}$ (Sect.4.3).

%%%%%%%%%%%%%%%%%%%%%%%%%%%%%%

Fig.\ref{grid2}b  shows the systematic pattern
after elimination of images with abnormally good and bad seeing
and illustrates the decrease of  systematic errors  compared to the 
use of a complete set of images (Fig.\ref{grid2}a). Few extrema
have vanished and most contrast details are smoothed. These
graphs correspond to the middle of the January 2007 epoch with the 
largest variance of monthly residuals (Fig.\ref{dxdyAA}),
which was chosen  to show
the worst case of $A(x,y)$. In November 2006 (first epoch)
these errors are much better (Fig.\ref{grid2}e).
The error structure for the merged $\langle V\rangle $ residuals
(Fig.\ref{grid2}d) in general is like that for $V_e$ ( Fig.\ref{grid2}b)
at the same epoch.  Fig.\ref{grid2}f refers to $V_e$ computed at
$k=6$ and  $R_{\rm {opt}}=0.5$\arcmin. The distribution shown here is similar  
to that for the reduction with $k=10$ (Fig.\ref{grid2}b) and for
$\langle V\rangle $ with  $\pm$50~$\mu$as  systematic errors at the center,
but much larger peaks at frame boundaries.

The dependence of $A(x,y)$ on $x,y$ could be 
a reason for excluding  systematic errors 
in a  secondary iteration;
however we had an
insufficient number of reference stars.
Alternatively,  $A_e$ can be  treated as additional 
components of each star model parameter $\xi$ to be computed 
with other parameters. 
This however is not useful for planet search or microlensing
applications due to the complete zeroing of useful astrometric signals.

Considering that a change of $A_e$ between two adjacent 
epochs at the frame center 
is approximately 50~$\mu$as (100--200~$\mu$as within the entire FoV) 
for both cameras,
we can estimate the stability of the FORS astrometric system over short
time scales. Assuming that
a given change occurs at about a month spacing, the daily rate of 
systematic change is  1--2~$\mu$as (3--6~$\mu$as for the  whole FoV).  
Our previous study of FORS1 errors
(Lazorenko et al. \cite{Lazorenko7}) have shown that
a difference in the systematic component 
over a time scale of four days is either undetectably small or 
30~$\mu$as atmost. This is about 8~$\mu$as a day 
change in  systematic errors 
in the whole FoV, which is in accord with the current estimate.

Recall that the amplitude of  systematic errors 
of 50~$\mu$as  we referred to corresponds to poor
observing conditions;  the estimate of
25~$\mu$as is more relevant for normal conditions and 
targets at frame center.

%%%%%%%%%%%%%%%%%%%%%%%%

\subsection{Instrumental background of  systematic errors}

Based on the discussion
in the previous subsection, we conclude
that the characteristics of the systematic component $A_e$ in magnitude 
and in spatial behavior are identical for FORS1 and FORS2. 
This is the  second  identity of these cameras
derived based on observations at rather different time scales,
CCD type, and photometric bands (the first is a dependence 
of image motion variance on $R$),
and is  evidently due to identical optical design of the cameras.

Systematic errors are probably generated during  
image centroiding due to the highly complicated
star profiles, actually indefinite at
the high accuracies at which we work. Consider that 
a typical error of 100~$\mu$as is only $2\times 10^{-4}$ fraction 
of the FWHM. Although FORS images are appropriate, the definition of the
'image photocenter' for star profiles distorted by 
variable optical aberrations becomes uncertain. Therefore 
we use a centroiding procedure (Lazorenko \cite{Lazorenko6}) 
specialized for finding the 
{\it weighted photocenter} of the image which is more
stable to image deformations than the "profile center".

At zero or constant geometric distortion, the 
position of the measured image centroid is subject only to  
random errors caused primarily by Poisson noise in the number of photons 
and by atmospheric image motion.
Systematic and extra random components in position appear in the following
cases of instability:

-- Small deformation of a star profile  (change of asymmetry) in time
due to a slow change of optical aberrations. This results in image 
photocenter shift proportional to the gradient of aberrations at the point.
These shifts,  correlated in $x,y$ space,  are
detected (after a certain filtration introduced
by astrometric reduction) as systematic residuals $A(x,y)$;
their change in time produces $A$ 
component. The effect is lowest at frame
center were images are most symmetric due to optimal optical performance.
 
-- A random change of image-to-image atmospheric PSF and seeing 
(shape and size). This
affects image profiles both in size and shape 
and therefore shifts photocenters by an amount  proportional
to seeing fluctuation {\it and} to the measure of 
image asymmetry at the point,  representing
modulation of  the optical aberration field by a random signal. 
The observed effect is
random in  time and correlated in space, 
and thus mimics atmospheric image motion.
In Sect.5 we classified it as "instrumental"
image motion with an amplitude exceeding that of atmospheric image motion and
not dependent on exposure time.

The combined effect of image asymmetry and seeing variation
is observed in an exaggerated form for stars with nearby companions,
the light of which  causes image asymmetry.
It is easily detected as a linear dependence of $V_{im}$ on FWHM,
which in most bad seeing conditions produces enormous systematic deviations 
exceeding by much the random errors and being  grounds for discarding 
these measurements (Sect.2).

Systematic deviations of the average seeing at a single epoch 
from its average  for a full set of data may produce systematic bias
of positions. 
To see whether this exists in our data, we performed
reduction with an expanded set of parameters ${\vec \xi}$
adding two extra terms describing the linear dependence of a star's $x,y$
position on seeing. With a full set (all seeing) of images, a new value
of the systematic error $A$ in an $r<1$\arcmin area
decreased from its former 160~$\mu$as to 90~$\mu$as, and a much 
smoother shape of $A(x,y)$ (Fig.\ref{grid}c)
was obtained in comparison to the initial pattern (Fig.\ref{grid}a).
The improvement, however, is seen primarily at the periphery
where geometric distortions are largest.
Thus variations of seeing actually lead to systematic errors in
epoch positions. We however consider the use of the expanded  reduction model
insufficiently validated, since it reduces the useful astrometric signal and,
for centrally placed targets, does not leads to a
significant improvement in the accuracy.
                        
Thus, variations in seeing are a source of systematic errors; for precise
astrometry these variations should be limited.
 
%\newpage
%tt 
%\newpage

\section{Allan precision}
\begin{figure}[htb]
\resizebox{\hsize}{!}{\includegraphics*[bb = 52 48 278 247 ]{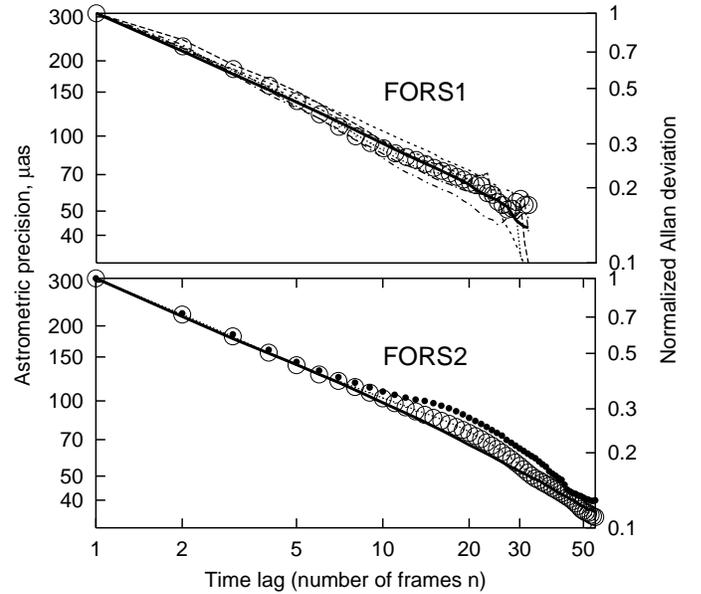}}
\caption {Normalized Allan deviation (right axis) 
in positional residuals 
and astrometric precision for bright targets
(left axis) expected from a series of $n$ images with standard seeing only.
Estimates are based on: - residuals $V$ obtained with 
$k=6...16$ (dashed lines  which for FORS2 
actually run into a single line); - merged residuals $\langle V\rangle $ 
(open circles); 
- numerical simulation assuming zero  systematic errors
(thick line); -  FORS2 residuals $V_m$  with 
no restriction on seeing (black dots). 
}
\label{allan}
\end{figure}

Another  characterization of 
the VLT long-term astrometric stability 
is based on the computation of the Allan deviation of residuals $V_m$. 
This quantity is normally used as a powerful indicator of 
systematic errors in observations and  corresponds
to the astrometric precision of a time series of $n$ images.
Fig.\ref{allan}  presents
plots of this variable as a function of time lag (expressed as
the number of images $n$) between
subsamples of   residuals $V_m$. 
The data in Fig.\ref{allan} are the average of each star Allan deviation
taken
over all, except the most peripheral, stars. Before averaging, we
normalized the individual Allan deviations to their values at zero time lag so as
to compensate for the highly varying amplitude 
of this variable for stars of different brightness. 
The right vertical axis of Fig.\ref{allan}
is the  scale for the normalized Allan deviation 
computed as described  and is valid for stars of any magnitude.
The left axis is the  scale used to find the Allan deviation
for bright stars ($B$=18 mag at FORS1 and $R$=16.5 mag at FORS2) 
with $\varepsilon_{}=230$~$\mu$as, $\delta=150$~$\mu$as  and
assuming its location  at the frame center, which ensures small
$\Delta_{\rm rf}=150$~$\mu$as.

%%%%%%%%%%%%%%%%%%%%%%%%%%%%%%

For computations, we used $V_m$ residuals obtained with each 
reduction parameter $k=6...16$  and  $R=R_{\rm {opt}}$. 
Results for different $k$ are very similar 
due to the  high degree of correlation between these data sets (dashed curves 
in Fig.\ref{allan}); 
for FORS2 they actually run into a single line corresponding to
the Allan deviation of residuals  $\langle V\rangle $.
Besides, 
due to the normalizing procedure, computations based on a subset of either all or 
bright stars only produced similar estimates. 
The Allan deviation was found to follow 
$n^{-0.522 \pm 0.002}$ (FORS1) and  $n^{-0.506 \pm 0.004}$ (FORS2)
power laws which are near to that
expected for the average of a random variable. A simple comparison
with an  $n^{-1/2}$ law is of course incorrect since least square fit residuals  
are correlated and have a non-diagonal covariance matrix $\vec{B}$. 
To compare our results with those expected with zero 
 systematic errors,  
we performed a numerical simulation of observations introducing
uncorrelated random noise in model measurements. The obtained dependence
(solid curves in Fig.\ref{allan}) follow a power law with a 
slope  $n^{-0.522 \pm 0.004}$ for both cameras, 
which is near to that obtained from observations.

The  difference in the observed and expected plots 
is clearly seen for the FORS2 camera starting from $n>15$.
This divergence is related to systematic errors discussed in Sect.6.
Due to the small magnitude of the errors,
they cause only a 5--10\% increase in the astrometric error 
in comparison to that expected in the absence of  systematic errors.

Above, in the case of  FORS2,  we used images with a standard FWHM 
to avoid any degradation of precision caused by
images with abnormal seeing.   
The Allan deviations with  all available measurements used
(black dots in Fig.\ref{allan}) are seen to have an excess 
of about 30\%  for  $n>15$.

\begin{figure}[htb]
\resizebox{\hsize}{!}{\includegraphics*[bb = 52 48 247 172 ]{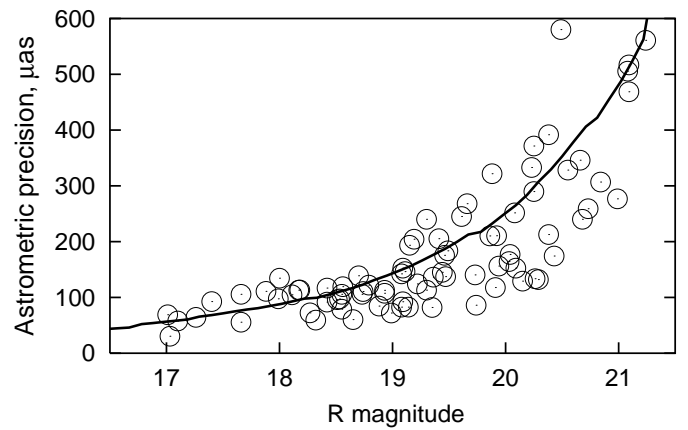}}
\caption { Astrometric precision (Allan deviation) 
for a series of 30 images as a function of  magnitude derived from
five-month  FORS2 observations 
(open circles); the same  expected in the case
of a very dense reference frame ($\Delta_{\rm rf} =0$) and 
moderate  $\delta = 150$~$\mu$as image motion (solid curve).
}                                  
\label{alac}
\end{figure}
%%%%%%%%%%%%%%%%%%%%%%%%%%%%%%

Given a series of 30 images (0.5--1 hours of telescope time), 
the precision of FORS1/2 astrometry
(Fig.\ref{allan}) is  about 50~$\mu$as.
At this
fixed number of images,  Fig.\ref{alac} (open circles)
shows the dependence of the precision on star brightness
computed for each star as the individual Allan deviation. Data refer
to FORS2 for which sufficiently long observation series 
are available and for images obtained at normal seeing. 
Because precision very much depends on
the reference frame noise (especially for bright targets),
the plot shows only those stars for which 
$\Delta_{\rm rf} \leq 300$~$\mu$as. For that reason, many
of the brightest 16.5--17.0 mag stars were omitted due to their peripheral
location in the frame.
The observed dependence of  precision on magnitude shows it to be
better than 100~$\mu$as for  $R<$19~mag targets and about 50~$\mu$as
for $R\approx 17$~mag.

The above estimate includes the error caused by reference frame and optical aberrations
which depend on a particular density of star distribution in the sky,
stability of the optical system, and
variations of seeing conditions at the period of observations.
Precision  improves with improved conditions.
A case of interest is the precision expected 
at high reference star density ($\Delta_{\rm rf} = 0$) 
and image motion $\delta = 150$~$\mu$as,  typical for a
reference frame size of 0.7--1.5\arcmin. These estimates, 
shown by a solid line in 
Fig.\ref{alac}, prove the feasibility of 50~$\mu$as astrometric precision
for brightest targets.

\section{Conclusion}

Astrometric quantities (residuals of positions, model parameters) derived 
from the processing of images are intrinsically relative. They are computed 
in a certain system and relative to the frame 
of reference objects  specific
to the particular target. These peculiarities of differential reduction
should be taken into account for the interpretation of the output data and 
in more complicated cases of handling  inhomogeneous series of images,
for instance obtained in different spectral bands or even with different
cameras, which is a case expected for long-term programmes. We demonstrated
that a careful processing of $B$ and $R$ images in a common system does not
degrade precision.

The precision of astrometric imaging at VLT
depends on several noise sources. Uncertainty of the
image photocenter determination 
$\varepsilon$, of course, is the dominant component of
the total error. Our data show that Eq.(\ref{eq:phsigma}) provides
a correct estimate of this error, at least with an accuracy of $\pm5$\%,
for a wide range of light fluxes and seeing conditions. 
$\varepsilon$ depends  not on the star magnitude  but 
on the light flux collected in the star image.
For that reason, the lowest $\varepsilon$ (equally, best astrometric
precision) is expected for images at saturation level, which depends on the
exposure, filter, pixel scale, and seeing. Therefore, having, for example,
dependence $\varepsilon(R)$ of $\varepsilon$ on magnitude $R$, it is easy
to apply it to other observations. For instance, in $B$ band 
$\varepsilon(B) \approx \varepsilon(R+\Delta m)$ where $\Delta m$ is
the difference of magnitudes  of images with equal light flux in
$B$ and $R$ bands. This scaling is illustrated in Fig.\ref{sig12} 
where $\varepsilon(B)$ and $\varepsilon(R)$ are seen to be 
the same dependences shifted by $\Delta m \approx 1.5$ mag.

Mitigation of atmospheric image motion at $T \geq 70$ s exposure does not 
present a problem due to its small amplitude in comparison to other random 
noise components. Quite unexpectedly, however, we found that the measured
image motion variance is the same at the very different $T=70$ and $T \approx 600$ s
exposure. We consider this as a new type of random error
caused by the  combined effect of the telescope-related
asymmetry of star profiles and of random
changes in the atmospheric PSF. We have found that images with abnormally
bad and good seeing are affected by large random and systematic errors
and thus their use should be avoided in precision astrometry.

Estimates of astrometric errors obtained in this study refer 
rather to {\it precision}
caused by stochastic error components but not to {\it accuracy}. 
The final astrometric accuracy can degrade compared to precision 
due to systematic errors incorporated into the reference frame
and then propagated by  the reduction model.
Presently, our astrometric results cannot be cross-compared against
independent datasets and analyses of common objects, which
could provide direct estimates of the long-term accuracy. 
In Sect.6,7, however, we demonstrated that 
the measured star displacements in space and time are 
fitted to nearly white noise residuals with a variance 
predicted by the model, therefore, an extra systematic component
(if present) should closely trace both the parallactic and proper motion of 
each star. A too low probability of this scenario 
(considering large accumulated Dof) implies
that the accuracy of the VLT relative parallax and proper motion
determination is comparable to the precision.

We have demonstrated that, with reference to systematic errors,
astrometry at FORS1/2 is accurate to 25~$\mu$as at five month
and 6 year time intervals. Due to that fact,
relative
proper motions and trigonometric parallaxes of stars in the test
field were derived
with a precision of 20~$\mu$as yr$^{-1}$ and 40~$\mu$as respectively
for 17--19 mag stars. Thus distances to stars at 1 kpc
can be measured with a precision of 4\%,
providing a correction from relative to absolute parallax is added.
This level of accuracy and, especially, good long-term 
stability, has numerous astrometric applications 
including  measurement of astrometric microlensing, 
planet detection and characterization by measuring reflex
motion of the parent star, and kinematics of Galactic stellar populations.
In the context of exoplanet searches, the use of the VLT
to search for planets  near brown dwarfs is very efficient.
With a 2 year observation programm, 
Saturn mass planets with orbit periods longer than 1 year 
and  Netpune mass planets with  2 year periods 
are detectable for brown dwarfs located at 10--20 pc. These
observations will make a real breakthrough in our understanding of planet
and brown dwarf formation by
probing a separation and mass range that is poorly suited to
other techniques.

Currently our method has several crucial limitations.
It is applicable to moderately populated sky areas with
low rates of blending but that are rich enough to  provide a
sufficient number of reference stars.
Also, targets brighter than 15-16 mag not can be 
measured due to saturation if the exposure time is not too short.
Therefore it is not  applicable to highly crowded sky areas like those
used for microlensing works and to bright, nearby solar-type stars
which are objects of interest for planet searches. Most of these 
limitations originate from the star profile fitting
and can be removed with use of a better  technique for determination of 
star photocenters.
Precision astrometry
is also problematic for telescopes with segmented primary mirrors that produce
intrinsic complex and time-variable PSF. 

We emphasize that the results of this study do not refer to 
a specific telescope, they validate precision astrometry for a whole class
of large ground-based imaging telescopes. Astrometry benefits 
highly from the use of large apertures, mitigating in this way both 
the principal image centroiding component 
of the total error and atmospheric image motion. For future 30 m telescopes,
the astrometric precision is expected to be 10$\mu$as or better for image
series of 0.5 hour duration.
At this precision, we are challenged with a wide spectrum of problems,
e.g.  the  ability to measure bright objects due to saturation
and the predominance of systematic errors caused by optical aberrations.

%\bibliograhystyle{aa}
%\bibliograhy{fors_bib.tex}

%\begin{acknowledgements}
%\end{acknowledgements}

%\bibliograhystyle{aa}

\end{document}